\documentclass[twocolumn]{autart}

\usepackage{bm}
\usepackage{enumitem}
\usepackage{commath}
\usepackage{graphicx}
\graphicspath{{Pictures/}}
\usepackage{amsmath}
\usepackage{subcaption}
\usepackage{breqn}
\usepackage{cite}
\usepackage[table]{xcolor}
\usepackage{booktabs}
\usepackage{empheq}
\usepackage{amsfonts}
\usepackage{amssymb}
\usepackage{hyperref}
\usepackage{xcolor}
\usepackage{mathrsfs}
\usepackage{accents}
\usepackage{float}
\usepackage{comment}
\usepackage{algpseudocode}
\usepackage{algorithm}

\newtheorem{lemo}{Lemma}
\newtheorem{defo}{Definition}
\newtheorem{asso}{Assumption}
\newtheorem{property}{Property}
\newtheorem{coro}{Corollary}

\newtheorem{remo}{Remark}
\newtheorem{thmo}{Theorem}

\newcommand{\rev}[1]{{\color{black}#1}}

\newcommand{\ubar}[1]{\underaccent{\bar}{#1}}

\newcommand{\myvar}[1]{\bm{#1}}
\newcommand{\tildevar}[1]{\tilde{\bm{#1}}}
\newcommand{\myvarfrak}[1]{\bm{\mathfrak{#1}}}

\newcommand{\myvardot}[1]{\dot{\myvar{#1}}}

\newcommand{\myvarddot}[1]{\ddot{\myvar{#1}}}

\newcommand{\myset}[1]{\mathcal{#1}}

\newcommand{\mysetbound}[1]{\partial \mathcal{#1}}
\newcommand{\mysetint}[1]{\mathring{\myset{#1}}}

\newcommand{\myub}[1]{\bar{#1}}
\newcommand{\mylb}[1]{\ubar{#1}}
\newcommand{\myubi}[1]{\myub{#1}_i}
\newcommand{\mylbi}[1]{\mylb{#1}_i}

\newcommand{\myubdot}[1]{\dot{\bar{#1}}}
\newcommand{\mylbdot}[1]{\dot{\ubar{#1}}}
\newcommand{\myubdoti}[1]{\myubdot{#1}_i}
\newcommand{\mylbdoti}[1]{\mylbdot{#1}_i}

\newcommand{\RN}[1]{\text{\MakeUppercase{\romannumeral #1}}}

\begin{document}

\begin{frontmatter}

\title{
Safe-by-Design Control for Euler-Lagrange Systems
}


\author{Wenceslao Shaw Cortez}\ead{wencsc@kth.se},     
\author{Dimos V. Dimarogonas}\ead{dimos@kth.se}               

\thanks[footnoteinfo]{This work was supported by the Swedish Research Council (VR),
the Swedish Foundation for Strategic Research (SSF), the Knut and Alice
Wallenberg Foundation (KAW) and the H2020-EU Research  and Innovation Programme under the GA No. 101016906 (CANOPIES).
The authors are with the School of EECS, Royal Institute of Technology (KTH), 100 44 Stockholm, Sweden.}

\begin{abstract}
Safety-critical control is characterized as ensuring constraint satisfaction for a given dynamical system. \rev{Recent developments in zeroing control barrier functions (ZCBFs) have provided a framework for ensuring safety of a superlevel set of a single constraint function. Euler-Lagrange systems represent many real-world systems including robots and vehicles, which must abide by safety-regulations, especially for use in human-occupied environments. These safety regulations include state constraints (position \emph{and} velocity) and input constraints that must be respected at all times. ZCBFs are valuable for satisfying system constraints for general nonlinear systems, however their construction to satisfy state \textit{and} input constraints is not straightforward. Furthermore, the existing barrier function methods do not address the multiple state constraints that are required for safety of Euler-Lagrange systems.} In this paper, we propose a methodology to construct multiple, non-conflicting control barrier functions for Euler-Lagrange systems subject to input constraints to satisfy safety regulations, while concurrently taking into account robustness margins and sampling-time effects. The proposed approach consists of \rev{a sampled-data controller} and an algorithm for barrier function construction to enforce safety (i.e satisfy position and velocity constraints). The proposed method is validated in simulation on a 2-DOF planar manipulator.

\end{abstract}

\end{frontmatter}

\section{Introduction}
Recent technological advancements have increased the presence of autonomous systems in human settings. The push for self-driving cars, drone delivery systems, and automated warehouses are a few examples of how autonomous systems are being exploited to improve efficiency and productivity. However, safety is key to properly incorporate these systems, particularly in human settings. The control of these autonomous systems must be able to guarantee safety of both the device and humans. 
 
\rev{Here we are motivated by safety in terms of the regulations provided by the International Standards Organization (ISO), which aim to ensure that machines respect position, velocity, and input constraints, e.g., do not leave a pre-defined region, exceed this speed, or apply excessive force, which must be respected at all times \cite{isosafety2011}}. Simultaneous satisfaction of these position, velocity, and input constraints for Euler-Lagrange systems renders the system \textbf{safe}. Furthermore, these systems are almost always controlled digitally in a sampled-data fashion and are prone to model uncertainties or external disturbances that must be accounted for. The problem addressed here is how to simultaneously satisfy input and system constraints for Euler-Lagrange systems to ensure \textbf{safety}.
 
Control barrier functions have attracted attention for constraint satisfaction of nonlinear systems. Existing barrier function methods have been applied to general nonlinear continuous/hybrid systems \cite{Prajna2007a} and used in control to satisfy constraints while providing stability \cite{Tee2009}. Those methods have been extended to less restrictive barrier function definitions and have been applied to bi-pedal walking, adaptive cruise control, and robotics \cite{Ames2014, Hsu2015, Rauscher2016, ShawCortez2018a}. Similar approaches have also addressed high relative degree systems \cite{Nguyen2016} and systems evolving on manifolds \cite{Wu2015a}. Recently, the distinction between reciprocal control barrier functions (RCBFs) and zeroing control barrier functions (ZCBFs) has been established \cite{Ames2017}, in which RCBFs are undefined at the constraint boundary while ZCBFs are zero at the boundary and well-defined outside of the constraint set. Aside from practical implementations, ZCBFs are advantageous in that they hold robustness properties in the form of input-to-state stability \cite{Xu2015a}. A review of existing approaches can be found in \cite{Ames2019}. 

\rev{Despite the novel developments in safety-critical methods using ZCBFs, no existing approach can ensure safety of Euler-Lagrange systems}. Recall that here we consider safety to incorporate state (position \emph{and} velocity) and input constraints. \rev{One existing method to handle state and input constraints} includes sum-of-squares programming \cite{Ames2019, Xu2018,Prajna2007a}, however that approach is only applicable to polynomial systems, and not to the Euler-Lagrange systems considered here. \rev{Another approach for addressing general state and input constraints} requires a pre-defined \rev{function (referred to as an evasive maneouver)} to then construct the ZCBF \cite{Squires2018, Ames2019}. However the design of the evasive maneouver is not straightforward in general, particularly with dynamically coupled systems such as Euler-Lagrange systems. 

\rev{Furthermore, a significant setback in existing ZCBF methods is their inability to couple position and velocity constraints, which is crucial for ensuring safety of the overall system. More specifically, the system should have bounded velocities and slow down as it approaches the boundary of the position constraint set. The existing approaches that address high relative degree \cite{Nguyen2016, Wu2015a} are prone to singularities in which the velocity is allowed to go unbounded inside the position workspace. These singularities occur when the gradient of the position constraint function is zero inside of the safe set, which prevents bounding the velocity even in standard norm-ball type position constraints (see \cite{Barbosa2020}). Recently, we developed methods to ``remove the singularity" in the high order barrier construction \cite{Tan2021}, however that approach only addresses the control input and not the velocity requirement.

 To address the singularity issue, here we consider multiple position constraints (e.g., box constraints), which collectively bounds the position without suffering from singularities and generalizes existing methods to handle multiple constraints simultaneously. The initial idea was presented in \cite{ShawCortez2019} where box constraints in the form of multiple ZCBFs were shown to naturally bound the velocity of the system and for which no such singularities occur. However addressing multiple ZCBFs while simultaneously handling input constraints requires ensuring that the multiple constraints are non-conflicting and has received little attention in the literature. In \cite{ShawCortez2019}, multiple ZCBFs were handled with input constraints in a sampled-data control law, however that method assumed that the controller was feasible. Other existing work has addressed multiple ZCBFs, but cannot handle input constraints \cite{Xu2018}. Recently, integral control barrier functions have been proposed as a means to satisfy input constraints \cite{Ames2021}, however for multiple ZCBFs, there is no guarantee that such a feasible control exists (see  Remark 4 of \cite{Ames2021}). Finally, recent methods using energy-based barrier functions are in fact able to simultaneously bound the position and velocity using a single barrier function, but insofar those methods cannot handle multiple barrier function nor input constraints \cite{ShawCortez2021,Singletary2020}. Thus despite the advances in safety-critical control, existing methods have yet to provide truly safe controllers for Euler-Lagrange systems.}  

In this paper, we present a methodology to construct ZCBFs for Euler-Lagrange systems. The proposed approach satisfies multiple workspace constraints (position \textit{and} velocity)\rev{, while simultaneously handling input constraints} to ensure safety of real-world systems. A correct-by-design algorithm is presented for the ZCBF construction that ensures forward invariance of the safe set, which can be computed off-line. The method considers robustness margins and sampling time effects. \rev{The main results are proposed for handling constraints in the form of box constraints, however we demonstrate how the approach hand be extended to more general constraint types}. The proposed approach is validated in numerical simulation on the 2-DOF planar manipulator. All of the code, including the algorithm to construct the the ZCBFs, is provided in \cite{ShawCortezCode}. A preliminary version of this work can be found in \cite{ShawCortez2020}. The approach presented here is less conservative than that of \cite{ShawCortez2020} and also relaxes the assumptions of \cite{ShawCortez2020}. Furthermore, the approach presented here also addresses robustness and sampling terms in the ZCBF construction, which are not considered in \cite{ShawCortez2020}.

\textit{Notation}: Throughout this paper, the term $\myvar{e}_j \in \mathbb{R}^{r}$ denotes the $j$th column of the identity matrix $I_{r\times r}$. The Lie derivatives of a function $h(\myvar{x})$ for the system $\myvardot{x} = \myvarfrak{f}(\myvar{x}) + \myvarfrak{g}(\myvar{x}) \myvar{u}$ are denoted by $L_{\mathfrak{f}} h = \frac{\partial h}{\partial x} \mathfrak{f}(\myvar{x})$ and $L_{\mathfrak{g}} h = \frac{\partial h}{\partial x} \myvar{g}(\myvar{x})$, respectively. The terms $\preceq$ and $\succeq$ are used to denote element-wise vector inequalities. The matrix inequality $A<B$ for square matrices $A$ and $B$ means that the matrix $B-A$ is positive-definite. The interior and boundary of a set $\myset{A}$ are denoted $\mysetint{A}$ and $\partial \myset{A}$, respectively. The notation $\alpha \circ \beta$ for a function $\alpha$ represents the composition $\alpha(\beta)$. We use the notation $x \searrow a$ and $x \nearrow a$, for some $a \in \mathbb{R}$, to denote the limit as $x$ approaches $a$ from above and below, respectively. \rev{A set $\myset{N}_p$ for $p \in \mathbb{N}$ is $\myset{N}_p = \{1,...,p\}$ .}

\section{Background}

\subsection{Control Barrier Functions}
Here we introduce the existing work regarding ZCBFs for nonlinear affine systems: $ \myvardot{x} = \myvarfrak{f}(\myvar{x}) + \myvarfrak{g}(\myvar{x}) \myvar{u}$, where $\myvar{x}(t) \in \mathbb{R}^n$ is the state, $\myvar{u} \in \mathbb{R}^m$ is the control input, $\myvarfrak{f}: \mathbb{R}^n \to \mathbb{R}^n$ and $\myvarfrak{g}: \mathbb{R}^n \to \mathbb{R}^{n\times m}$ are locally Lipschitz continuous. We denote $\myset{I}\subseteq \mathbb{R}_{\geq 0}$, where $0 \in \myset{I}$, as the maximal interval of existence of $\myvar{x}(t)$. A set $\myset{S} \subset \mathbb{R}^n$ is forward invariant if $\myvar{x}(0) \in \myset{S}$ implies $\myvar{x}(t) \in \myset{S}$ for all $t \in \myset{I}$.

Let $h(\myvar{x}): \mathbb{R}^n \to \mathbb{R}$ be a continuously differentiable function, and let the associated constraint set be defined by:
\begin{align}\label{eq:constraint set general}
\myset{C} = \{\myvar{x} \in \mathbb{R}^n: h(\myvar{x}) \geq 0\}
\end{align}
The function $h$ is considered the zeroing control barrier function and formerly defined as:
\begin{defo}[\cite{Ames2019}] \label{def:zcbf}
Let $\myset{C} \subset \myset{E} \subset \mathbb{R}^n$ defined by \eqref{eq:constraint set general} be the superlevel set of a continuously differentiable function $h: \myset{E} \to \mathbb{R}$, then $h$ is a zeroing control barrier function if there exists an extended class-$\mathcal{K}_\infty$ function $\alpha$ such that for the control system $\myvardot{x} = \myvarfrak{f}(\myvar{x}) + \myvarfrak{g}(\myvar{x}) \myvar{u}$, the following holds: $
\underset{\myvar{u} \in \myset{U}}{\text{sup}} [L_{\mathfrak{f}} h (\myvar{x}) + L_{\mathfrak{g}} h(\myvar{x}) \myvar{u} ] \geq - \alpha(h(\myvar{x})), \forall \myvar{x} \in \myset{E}$
\end{defo}

If $h$ is a zeroing control barrier function, the condition $\dot{h}(\myvar{x}) \geq -\alpha(h(\myvar{x}))$ is then enforced in the control by re-writing it as: $L_{\mathfrak{f}} h + L_{\mathfrak{g}} h \myvar{u} \geq -\alpha(h(\myvar{x}))$, which is linear with respect to $\myvar{u}$, and ensures forward invariance of $\myset{C}$ \cite{Ames2019}. We further note that the ZCBF conditions can be extended to sampled-data systems for which $\myvar{u}$ is piece-wise continuous. That is, for a ZCBF $h$ where $\dot{h}(\myvar{x}) \geq -\alpha(h(\myvar{x}))$ holds for almost all $t \in [0, T]$ \rev{(see \cite{ShawCortez2019, Glotfelter2017})}.

\subsection{\rev{System} Dynamics}
Consider the following \rev{dynamical} system for the generalized coordinates $\myvar{q}, \myvar{v} \in \mathbb{R}^n$:
\begin{equation}\label{eq:nonlinear affine dynamics}
\begin{split}
\myvardot{q} &= \myvar{v}  \\
\myvardot{v} &= \rev{G(\myvar{q})\left( \myvar{f}_1(\myvar{q}, \myvar{v}) + \myvar{f}_2(\myvar{q}, \myvar{v}) + \myvar{f}_3(\myvar{q})+ \myvar{u} \right) }
\end{split}
\end{equation}
where \rev{$G(\myvar{q}) \in \mathbb{R}^{n \times m}$ and $\myvar{f}_1(\myvar{q}, \myvar{v}), \myvar{f}_2(\myvar{q}, \myvar{v}), \myvar{f}_3(\myvar{q}) \in \mathbb{R}^m$ are globally Lipschitz continuous functions, and  $\myvar{u} \in \myset{U} \subset \mathbb{R}^m$ is the control input. }


Here we consider the following well-known properties for Euler-Lagrange systems \cite{Ortega1998}:
\begin{property}:\label{prop:M}
\rev{$G(\myvar{q})$ is full row rank such that there exists a $G^+(\myvar{q}) \in \mathbb{R}^{m\times n}$ for which $G G^+ = I_{n \times n}$.}
\end{property}
\begin{property}:\label{prop:C}
There exists $k_c \in \mathbb{R}_{>0}$ \rev{such that $\| \myvar{f}_1(\myvar{q}, \myvar{v}) \| \leq k_c \|\myvar{v}\|^2$,  $\forall (\myvar{q}, \myvar{v}) \in \mathbb{R}^n \times \mathbb{R}^n$}.
\end{property}
\begin{property}:\label{prop:F}
\rev{There exist constants $f_j \in \mathbb{R}_{\geq 0}$ such that $\| \myvar{e}_j^T f_2(\myvar{q},\myvar{v})\| \leq f_j \|\myvar{v}\|$, $\forall (\myvar{q}, \myvar{v}) \in \mathbb{R}^n \times \mathbb{R}^n$, $\forall j \in \myset{N}_m$.}
\end{property}

\begin{remo}
Note that Properties \ref{prop:M}-\ref{prop:F} and \rev{global Lipschitz continuity of the dynamics} can be relaxed and are only taken for simplicity here. \rev{We consider safety of the system for $(\myvar{q}, \myvar{v}) \in \myset{H}$ for a compact set $\myset{H}$ such that these properties can be checked in a compact set containing $\myset{H}$ instead of over the entire space $\mathbb{R}^n \times \mathbb{R}^n$}.
\end{remo}

\subsection{Problem Formulation}

The goal of constraint satisfaction is to ensure the states $\myvar{q}, \myvar{v}$ stay within a set of constraint-admissible states. Here we focus on workspace constraints reminiscent of real-world systems which are defined by:
\begin{equation}\label{eq:constraint set multiple position}
 \myset{Q} = \{ \myvar{q} \in \mathbb{R}^n: \myvar{q}_{min} \preceq \myvar{q} \preceq \myvar{q}_{max} \}
 \end{equation}
for $\myvar{q}_{min}, \myvar{q}_{max} \in \mathbb{R}^n$ and $\myvar{q}_{max} \succ \myvar{q}_{min}$. These types of constraints are highly applicable in robotics and general automated systems. 

We further address the velocity constraints that the system must satisfy as:
\begin{equation}\label{eq:constraint set multiple velocity}
\myset{V} = \{ \myvar{v} \in \mathbb{R}^n: \myvar{v}_{min} \preceq \myvar{v} \preceq \myvar{v}_{max} \}
\end{equation}
where $\myvar{v}_{min}, \myvar{v}_{max} \in \mathbb{R}^n$, $\myvar{v}_{max} \succ 0$, and for simplicity of the presentation let $\myvar{v}_{min} = -\myvar{v}_{max}$.

In addition to state constraints, real-world systems have limited actuation capabilities. Thus the aforementioned state constraints must be realizable with the available control inputs. Let $\myset{U}$ be the available control inputs:
\begin{equation}\label{eq:constraint set input}
\myset{U} = \{ \myvar{u} \in  \rev{\mathbb{R}^m}: \myvar{u}_{min} \preceq \myvar{u} \preceq \myvar{u}_{max} \}
\end{equation}
where $\myvar{u}_{min}, \myvar{u}_{max} \in \rev{\mathbb{R}^m}$, $\myvar{u}_{max} \succ 0$, and for simplicity of the presentation let $\myvar{u}_{min} = - \myvar{u}_{max}$.

The problem addressed here is to design a control law that renders the set of state constraints forward invariant. We formally define a \textit{safe} system as follows:
\begin{defo}\label{def:safe set}
Consider the constraint sets \eqref{eq:constraint set multiple position}, \eqref{eq:constraint set multiple velocity}, and \eqref{eq:constraint set input}. \rev{Suppose for a given $\myvar{u}$, \eqref{eq:nonlinear affine dynamics} with initial condition $(\myvar{q}(0), \myvar{v}(0)) \in \mathbb{R}^n \times \mathbb{R}^n$ admits an absolutely continuous solution $(\myvar{q}(t), \myvar{v}(t))$ for all $t \in \myset{I} \subset \mathbb{R}_{>0}$}. The system \eqref{eq:nonlinear affine dynamics} is considered \textbf{safe} if for any $(\myvar{q}(0), \myvar{v}(0)) \in \myset{Q} \times \myset{V}$, \rev{$\myset{I} \subseteq \mathbb{R}_{\geq 0}$ and }$(\myvar{q}(t), \myvar{v}(t)) \in \myset{Q} \times \myset{V}$ for all $t \geq 0$.
\end{defo}

We note that this definition of safety is stronger than forward invariance of the constraint set as we require forward invariance for all $t \geq 0$. The problem addressed here is formally stated as follows:
\begin{prob}\label{prob:main problem}
Consider the system \eqref{eq:nonlinear affine dynamics} with position, velocity, and input constraints \eqref{eq:constraint set multiple position}, \eqref{eq:constraint set multiple velocity}, \eqref{eq:constraint set input}. Design a control law $\myvar{u} \in \myset{U}$ that renders \eqref{eq:nonlinear affine dynamics} \textbf{safe}.
\end{prob}

\section{Proposed Solution}
In this section, we present the candidate ZCBFs and the proposed control laws to ensure safety. We first construct the candidate ZCBFs with design parameters. We proceed to construct bounds on the design parameters such that system safety is ensured under the condition that $\myvar{u} \in \myset{U}$. The construction of the design parameters yields an algorithm for constructing ZCBFs. Finally, we design continuous-time and sampled-data control laws to guarantee system safety.

\subsection{ZCBF Construction}
In this section, we construct the ZCBFs for system safety. We note that the construction is motivated by the approach from \cite{ShawCortez2019} \rev{(with similar high order barrier techniques as \cite{Wu2015a, Nguyen2016, Tan2021})}, although in a less conservative manner as will be discussed later. To define the ZCBFs, we re-write the constraint set $\myset{Q}$ into individual constraints with respect to functions $\myubi{h}$, $\mylbi{h}: \mathbb{R} \to \mathbb{R}$, which are defined as:
\begin{equation}\label{eq:zcbf h}
\myubi{h}(q_i) = q_{max_i} - q_i, \ \mylbi{h} = q_i - q_{min_i}, \ i\in \myset{N}_n
\end{equation}
where $q_{max_i}$, $q_{min_i} \in \mathbb{R}$ are the $i$th elements of $\myvar{q}_{max}, \myvar{q}_{min}$, respectively, from \eqref{eq:constraint set multiple position}. We define the superlevel set of $\myubi{h}$ and $\mylbi{h}$ as:
\begin{equation}\label{eq:constraint set Ci}
\myset{Q}_i = \{ q_i \in \mathbb{R}: \myubi{h}(q_i) \geq 0, \mylbi{h}(q_i) \geq 0\}, \ i \in \myset{N}_n
\end{equation}
\rev{It follows that:} $\myset{Q}_1 \times ... \times \myset{Q}_n = \myset{Q}$.

In order to define a superset of $\myset{Q}$ over which the ZCBF conditions hold, we introduce the following functions: 
\begin{equation}\label{eq:zcbf h delta}
\myubi{h}^\delta(q_i) = \myubi{h}(q_i) + \delta, \ \mylbi{h}^\delta(q_i) = \mylbi{h}(q_i) + \delta, \ i\in \myset{N}_n
\end{equation}
where $\delta \in \mathbb{R}_{\geq 0}$ is a design parameter. We similarly define a superlevel set for $\myubi{h}^\delta$ and $\mylbi{h}^\delta$ as:
\begin{equation}\label{eq:constraint set Ci delta}
\myset{Q}^\delta_i = \{ q_i \in \mathbb{R}: \myubi{h}^\delta(q_i) \geq 0, \mylbi{h}^\delta(q_i) \geq 0\}, \ i \in \myset{N}_n
\end{equation}
Note that $\myset{Q}_i \subset \myset{Q}_i^\delta$ for $\delta >0$ and $\myset{Q}_i = \myset{Q}_i^\delta$ if $\delta = 0$. \rev{Let} $\myset{Q}^\delta = \myset{Q}^\delta_1 \times ... \times \myset{Q}^\delta_n$, \rev{and we note that} $\myset{Q} = \myset{Q}^0$. Moreover, consideration of $\myset{Q}^\delta$ for $\delta > 0$ allows for consideration of robustness to perturbations in the proposed formulation. We refer to \cite{Xu2015a} for a discussion on robustness of ZCBFs.

We now introduce new functions to address the relative-degree of the system: $\myubi{b}$, $\mylbi{b}: \mathbb{R}\times \mathbb{R}\to \mathbb{R}$, and we treat these functions as the candidate ZCBFs for Euler-Lagrange systems defined as follows:
\begin{align}\label{eq:zcbf b}
&\myubi{b}(q_i,v_i) = -v_i + \gamma \alpha(\myubi{h}(q_i)), \notag \\
&\mylbi{b}(q_i,v_i) = v_i + \gamma \alpha(\mylbi{h}(q_i)), \hspace{.3cm} \ i \in \myset{N}_n
\end{align}
where $\alpha$ is a continuously differentiable, extended class-$\mathcal{K}_\infty$ function, and $\gamma \in \mathbb{R}_{>0}$ is a design parameter. We see that when $\myubi{b}\geq 0$ and $\mylbi{b}\geq 0$ it follows that $\myubdoti{h} \geq -\gamma\alpha(\myubi{h})$ $\mylbdoti{h} \geq -\gamma\alpha(\mylbi{h})$ as required from Definition \rev{\ref{def:zcbf}} for forward invariance of $\myset{Q}_i$. 

We treat $\myubi{b} \geq 0$ and $\mylbi{b}\geq 0$,  $\forall i \in \myset{N}_n$ as new constraints to be satisfied. To properly address the set of states where $\myubi{b}\geq 0$ and $\mylbi{b}\geq 0$, we define the following set:
\begin{equation}\label{eq:set B definition}
\myset{B}_i = \{ (q_i,v_i) \in \mathbb{R} \times \mathbb{R} : \myubi{b}(q_i,v_i )\geq 0, \mylbi{b}(q_i,v_i)\geq 0 \}
\end{equation}
with $\myset{B} = \myset{B}_1 \times ... \times \myset{B}_n$.

We define the following functions to define supersets of $\myset{B}$:
\begin{align}\label{eq:zcbf b delta}
&\myubi{b}^\delta(q_i,v_i) = -v_i + \gamma \alpha(\myubi{h}^\delta(q_i)), \notag \\
&\mylbi{b}^\delta(q_i,v_i) = v_i + \gamma \alpha(\mylbi{h}^\delta(q_i)), \hspace{.3cm} \ i \in \myset{N}_n,
\end{align}
with the following superlevel sets:
\begin{equation}\label{eq:set B definition delta}
\myset{B}^\delta_i = \{ (q_i,v_i) \in \mathbb{R} \times \mathbb{R} : \myubi{b}^\delta(q_i,v_i )\geq 0, \mylbi{b}^\delta(q_i,v_i)\geq 0 \}
\end{equation}
and $\myset{B}^\delta = \myset{B}^\delta_1 \times ... \times \myset{B}^\delta_n$. By construction, it follows that $\myset{B}^\delta \supset \myset{B}$ for $\delta > 0$ and $\myset{B}^0 = \myset{B}$.

Next we define the intersection of $\myset{Q}_i$ and $\myset{B}_i$ and the respective superset as:
\begin{equation}\label{eq:set H definition}
\myset{H}_i := (\myset{Q}_i \times \mathbb{R}) \cap \myset{B}_i, \ i\in \myset{N}_n
\end{equation}
\begin{equation}\label{eq:set H definition delta}
\myset{H}^\delta_i := (\myset{Q}^\delta_i \times \mathbb{R}) \cap \myset{B}^\delta_i, \ i\in \myset{N}_n
\end{equation}
with $\myset{H} = \myset{H}_1 \times ... \times \myset{H}_n$ and $\myset{H}^\delta = \myset{H}^\delta_1 \times ... \times \myset{H}^\delta_n$ such that $\myset{H}^\delta \supset \myset{H}$ if $\delta > 0$ and $\myset{H}^0 = \myset{H}$. We denote $\myset{H}$ as the \textit{safe set}. Note that since $\myset{Q}$ is compact, so are $\myset{H}_i$ and $\myset{H}_i^\delta$ for $i \in \myset{N}_n$.

In order to ensure forward invariance of $\myset{B}$ (and thus $\myset{Q}$), we repeat the ZCBF conditions as per \rev{Definition \ref{def:zcbf}} with respect to the ZCBF candidates $\myubi{b}, \mylbi{b}$:
\begin{align}\label{eq:Nagumo rel 2 condition}
& \myubdoti{b}(q_i,v_i) \geq - \nu \beta(\myubi{b}(q_i,v_i)) + \bar{\eta}, \notag \\
&\mylbdoti{b}(q_i,v_i) \geq - \nu \beta(\mylbi{b}(q_i,v_i)) + \bar{\eta}
\end{align}
for all $(q_i,v_i) \in \myset{H}^\delta$, $i \in \myset{N}_n$, where $\beta$ is an extended class-$\mathcal{K}_\infty$ function, $\nu \in \mathbb{R}_{>0}$ is an additional barrier function design parameter, and $\bar{\eta} \in \mathbb{R}_{\geq 0}$ is an added term motivated by \cite{ShawCortez2019} to incorporate sampling-time effects into the proposed ZCBF construction. We note that for $\bar{\eta} := 0$, \eqref{eq:Nagumo rel 2 condition} follows the conventional requirements for ZCBFs \cite{Ames2019}.

Substitution of \eqref{eq:nonlinear affine dynamics} into \eqref{eq:Nagumo rel 2 condition} and concatenation over all $i \in \myset{N}_n$ yields:
\begin{align}\label{eq:Nagumos condition matrix}
S \rev{G(\myvar{q}) ( \myvar{f}_1(\myvar{q}, \myvar{v}) \myvar{v} + \myvar{f}_2(\myvar{q}, \myvar{v}) + \myvar{f}_3(\myvar{q})} + \myvar{u} ) + \gamma \Lambda(\myvar{q}) S \myvar{v} \notag \\
 \succeq - \nu \myvar{p}(\myvar{q}, \myvar{v}) + \bar{\eta} \myvar{1}_{2n}
\end{align}
where $S$ $= $ $[-I_{n\times n}$ $, I_{n\times n}]^T$, $\Lambda(\myvar{q})$ $= $ $\text{diag}\{\frac{\partial \alpha}{\partial \myub{h}_1}(\myvar{q})$ $,... $ $\frac{\partial \alpha}{\partial \myub{h}_n}(\myvar{q})$ $, \frac{\partial \alpha}{\partial \mylb{h}_1}(\myvar{q})$ $, ...$ $,\frac{\partial \alpha}{\partial \mylb{h}_n}(\myvar{q}) \}$ and $\myvar{p}(\myvar{q}, \myvar{v})$ $:= [\beta \circ \myub{b}_1(q_1,v_1), ..., \beta \circ \myub{b}_n(q_n,v_n),\beta \circ \mylb{b}_1(q_1,v_1), ..., \beta \circ \mylb{b}_n(q_n,v_n) ]^T$. 

To summarize, satisfaction of \eqref{eq:Nagumos condition matrix} for all $(\myvar{q}, \myvar{v}) \in \myset{H}^\delta \supset \myset{H}$ for some $\delta > 0$ ensures \eqref{eq:Nagumo rel 2 condition} holds for all $(q_i, v_i) \in \myset{H}^\delta_i$, $i \in \myset{N}_n$, which in turn ensures $q_i \in \myset{Q}_i$ for all $i \in \myset{N}_n$.

We note that \eqref{eq:Nagumos condition matrix} is linear with respect to $\myvar{u}$, and define the proposed quadratic program-based control law:
\begin{align} \label{eq:consat proposed ct}
\begin{split}
\myvar{u}^*(\myvar{q}, \myvar{v}) \hspace{0.1cm} = \hspace{0.1cm} & \underset{\myvar{u} \in \myset{U}}{\text{argmin}}
\hspace{.3cm} \| \myvar{u} -\myvar{u}_{\text{nom}}(\myvar{q}, \myvar{v},\rev{t}) \|^2_2  \\
& \text{s.t.} \hspace{.9cm} \eqref{eq:Nagumos condition matrix}
\end{split}
\end{align}
where $\myvar{u}_{nom}: \mathbb{R}^n \times \mathbb{R}^n \rev{\times \mathbb{R}} \to \mathbb{R}^{n}$ is some nominal control law which can represent, for example, a pre-defined stabilizing controller or possibly a human input to the system. Implementation of \eqref{eq:consat proposed ct}, assuming a solution exists, can be used to ensure forward invariance of $\myset{H}$. \rev{We emphasize that the construction of the barriers leads to linearly \emph{dependent} input constraints as can be seen in \eqref{eq:consat proposed ct} due to $S$ and the input constraints. This is in contrast with the QP-based control formulations, typical of ZCBF controllers, which require linear independence of the constraints on $\myvar{u}$ \cite{MOrris2015, Hager1979}. For continuous-time implementations, linear independence of the QP constraints is important for ensuring local Lipschitz continuity of $\myvar{u}^*$. Here, we allow for discontinuous, sampled-data control, which is impartial to linear dependency in the control constraints since the control is inherently non-Lipschitz.}

Before we present the main theorem, we must state two assumptions to be satisfied. First, we make the following realistic assumption that the system has sufficient control authority in the set $\myset{Q}^\delta$:
\begin{asso}\label{asm:control}
There is sufficient control authority such that for given $\delta, \bar{\eta} \in \mathbb{R}_{\geq 0}$,
there exists some $\varepsilon \in \mathbb{R}_{>0}$ such that \rev{$u_{\max_j} > |\myvar{e}_j^T \myvar{f}_3(\myvar{q})| +( \varepsilon + \bar{\eta}) \| \myvar{e}_j^T G(\myvar{q})^+ \|_\infty$ for all $\myvar{q}\in \myset{Q}^\delta$, $j \in \myset{N}_m$.}
\end{asso}
This is a common assumption to ensure that in fact the system can be held statically and has the capability to move from any configuration over $\myset{Q}^\delta$. From a pragmatic perspective, we note that this assumption is  always satisfied in practice in order for the system to perform a desired task. Furthermore, this assumption is much less conservative than that of \cite{ShawCortez2020}, which effectively requires each $u_i$ to satisfy \eqref{eq:Nagumo rel 2 condition} independently while all other inputs are at their respective maximum values (i.e., $|u_j| = u_{max_j}$ for all $j \neq i$).  

Second, we require the extended class-$\mathcal{K}_\infty$ functions $\alpha$ and $\beta$ to satisfy the following properties:
\begin{asso}\label{asm:beta}
Given a $\delta \in \mathbb{R}_{\geq 0}$, the extended class-$\mathcal{K}_\infty$ functions $\alpha$ and $\beta$ satisfy the following conditions:
\begin{enumerate}
\item There exists a $d \in \mathbb{R}_{>0}$ such that  $\alpha(-e) + \alpha(q_{max_i} - q_{min_i} + e) \geq d$ holds for all $i \in \myset{N}_n$, $e \in [0, \delta]$.
\item For any $a,c \in \mathbb{R}_{>0}$ and $b \in \mathbb{R}_{\geq 0}$ such that $a - b = 2c$, then $\beta$ satisfies: $\beta(a) + \beta(-b) \geq \beta(c)$.
\end{enumerate} 
\end{asso}
Assumption \ref{asm:beta} requires that the slope of $\alpha$ and $\beta$ on the negative real-axis is sufficiently small with respect to that of the positive real-axis. This condition is required to consider how the system behaves in $\myset{H}^\delta \setminus \myset{H}$. For example, if a disturbance exists that pushes the system into $\myset{H}^\delta \setminus \myset{H}$ (where $\myubi{b}<0$ or $\mylbi{b}<0$), the restoring ``force" that keeps the system ultimately bounded \cite{ShawCortez2019} must not exceed the capabilities of the actuators. 
\begin{remo}
Assumption \ref{asm:beta} is not restricted to linear functions used in ``exponential barrier functions" \cite{Nguyen2016}, nor polynomial functions used in sum-of-squares programming techniques \cite{Prajna2007a}. Assumption \ref{asm:beta} only restricts the slope of the two extended class-$\mathcal{K}_\infty$ functions over the negative real-axis. As a result of this generality, both linear functions and (odd) polynomial functions are subclasses of functions that satisfy Assumption \ref{asm:beta}. 
\end{remo}
In the following theorem, we ensure a solution to \eqref{eq:consat proposed ct} always exists for all $(\myvar{q}, \myvar{v}) \in \myset{H}^\delta$ by appropriately computing $\gamma$ and $\nu$:

\begin{thmo}\label{thm:existenceZCBF}
Consider the system \eqref{eq:nonlinear affine dynamics} with the state and input constraints defined by \eqref{eq:constraint set multiple position}, \eqref{eq:constraint set multiple velocity}, and \eqref{eq:constraint set input}. Let the set $\myset{H}^\delta_i$ be defined by \eqref{eq:set H definition delta} for $i \in \myset{N}_n$ with the continuously differentiable extended class-$\mathcal{K}_\infty$ function $\alpha $ and extended class-$\mathcal{K}_\infty$ function $\beta$. Suppose Assumptions \ref{asm:control} and \ref{asm:beta} hold for sufficiently small $\delta$ \footnote{We note that Definition \ref{def:zcbf} requires $\myset{E} \supset \myset{C}$ which, equivalently stated, requires $\delta > 0$. For the sake of generality we show that the results of Theorem \ref{thm:existenceZCBF} hold for $\delta = 0$, although in Section \ref{ssec:control implementation} we also require $\delta > 0$.}, $\bar{\eta} \in \mathbb{R}_{\geq0}$. Then there exist $\gamma_1^*$, $\gamma_2^*$, $\gamma_3^*$, $\nu_1^*$, $\nu_2^* \in \mathbb{R}_{>0}$  (with $\nu_1^* < \nu_2^*$) such that the choice of $\gamma \in (0, \min\{ \gamma_1^*, \gamma_2^*, \gamma_3^*\}]$, $\nu \in [\nu_1^*, \nu_2^*]$ if $\delta > 0$ otherwise $\nu \geq \nu_1^*$ if $\delta = 0$, ensures that $\myvar{u}^*$ defined by \eqref{eq:consat proposed ct} exists and is unique for all $(\myvar{q}, \myvar{v}) \in \myset{H}^\delta$. Furthermore, for any $(\myvar{q}, \myvar{v}) \in \myset{H}^\delta$, $\myvar{v} \in \myset{V}$.
\end{thmo}
The proof of Theorem \ref{thm:existenceZCBF} is constructive. In the following section, we analyze the properties of $\myubi{b}$ and $\mylbi{b}$ for Euler-Lagrange systems to construct $\gamma_1^*$, $\gamma_2^*$, $\gamma_3^*$, $\nu_1^*$, and $\nu_2^*$, as well as bounds on $\delta$, $\bar{\eta}$, and a valid control $\tildevar{u} \in \myset{U}$ such that there always exists a solution to \eqref{eq:consat proposed ct}.

\subsection{Analysis}\label{ssec:analysis}
In this section, we present properties of $\myset{H}^\delta$ in relation to the candidate ZCBFs $\myubi{b}$, and $\mylbi{b}$ to design $\gamma$ and $\nu$.
\subsubsection{\textbf{Velocity Relations}}
First, we state the following Lemma to relate the system velocity with $\myset{H}^\delta$:
\begin{lemo}\label{lem:velocity relations}
Consider the functions and sets $\myubi{h}^\delta$, $\mylbi{h}^\delta$, $\myubi{b}^\delta$, $\mylbi{b}^\delta$, $\myset{Q}^\delta_i$, $\myset{B}^\delta_i$, and $\myset{H}^\delta_i$ defined respectively by \eqref{eq:zcbf h delta}, \eqref{eq:zcbf b delta}, \eqref{eq:constraint set Ci delta}, \eqref{eq:set B definition delta}, \eqref{eq:set H definition delta} with extended class $\mathcal{K}_\infty$ function $\alpha$. Then for $\delta \geq 0$, $\|\myvar{v}\|_\infty \leq \bar{v}$ for all $(\myvar{q}, \myvar{v}) \in \myset{H}^\delta$, where
\begin{align}\label{eq:velocity bound}
\bar{v} = \gamma a:= \gamma \alpha (2\delta + \|\myvar{q}_{min} - \myvar{q}_{max}\|_\infty)
\end{align}
\end{lemo}
\begin{pf}
From \eqref{eq:constraint set Ci delta}, \eqref{eq:zcbf b delta}, and \eqref{eq:set B definition delta} it follows that $-\gamma \alpha(\mylbi{h}^\delta(q_i))$ $\leq v_i$ $\leq$ $\gamma \alpha(\myubi{h}^\delta(q_i))$ for $(q_i,v_i) \in \myset{H}^\delta$, $i \in \myset{N}_n$. Thus $\myvar{v}$ is bounded in $\myset{H}^\delta$. Furthermore, the maximum value of $\alpha(\myubi{h}^\delta(q_i))$ and $\alpha(\mylbi{h}^\delta(q_i))$ in $\myset{Q}_i^\delta$ for $i \in \myset{N}_n$ is $a = \alpha (2\delta + \|\myvar{q}_{min} - \myvar{q}_{max}\|_\infty)$, which yields $\gamma a$ as the maximum value of $\myvar{v}$ and completes the proof.
\end{pf}

Lemma \ref{lem:velocity relations} provides insight into how the ZCBF construction affects the system behaviour. First, by appropriately tuning $\gamma$, the velocity bounds from \eqref{eq:velocity bound} can be adjusted to satisfy the state constraint $\myvar{v} \in \myset{V}$. Second, the relation $-\gamma \alpha(\mylbi{h}(q_i)) \leq v_i \leq \gamma \alpha(\myubi{h}(q_i))$ shows that as $\myvar{q}$ approaches the boundary $\mysetbound{Q}$, the velocity approaches zero. This is an important property because it restricts the system's inertia relative to the constraint boundary. This aligns with intuition in that if the velocity is too high near the boundary, exceedingly large control effort would be required to ensure forward invariance. While $\gamma$ dictates the system's velocity, $\nu$ dictates the behaviour of $\myvar{u}$ as the system approaches the constraint boundary. From \eqref{eq:Nagumo rel 2 condition}, $\nu$ will dictate how soon the control acts to keep the system in the constraint set. 

From Lemma \ref{lem:velocity relations}, we define the following upper bound on $\gamma$ such that the maximum velocity will be contained in $\myset{V}$ to ensure safety: 
\begin{align}\label{eq:gamma star 3}
\gamma_1^* := \dfrac{1}{a} \underset{i \in \myset{N}_n}{\text{min}} v_{max_i}
\end{align}
where $a \in \mathbb{R}_{>0}$ is defined in \eqref{eq:velocity bound}.
\begin{lemo}\label{lem:velocity consat}
Consider the functions and sets $\myubi{h}^\delta$, $\mylbi{h}^\delta$, $\myubi{b}^\delta$, $\mylbi{b}^\delta$, $\myset{Q}^\delta_i$, $\myset{B}^\delta_i$, $\myset{H}^\delta_i$, and $\myset{V}$ defined respectively by \eqref{eq:zcbf h delta}, \eqref{eq:zcbf b delta}, \eqref{eq:constraint set Ci delta}, \eqref{eq:set B definition delta}, \eqref{eq:set H definition delta}, \eqref{eq:constraint set multiple velocity} with extended class $\mathcal{K}_\infty$ function $\alpha$. If $\delta \in \mathbb{R}_{\geq 0}$, then $\gamma_1^*$ defined by \eqref{eq:gamma star 3} is strictly positive, and if $\gamma \in (0, \gamma_1^*]$, then $\myvar{v} \in \myset{V}$ for all $(\myvar{q}, \myvar{v}) \in \myset{H}^\delta$.
\end{lemo}
\begin{pf}
Strict positivity of $\gamma_1^*$ follows since $v_{max_i} > 0$ and for $\delta \geq 0$, $a > 0$ due to $\myvar{q}_{max} \succ \myvar{q}_{min}$. From Lemma \ref{lem:velocity relations} it follows that $\|\myvar{v} \|_\infty \leq \gamma a$. To ensure $\myvar{v} \in \myset{V}$, i.e., $\myvar{v}_{min} \preceq \myvar{v} \preceq \myvar{v}_{max}$, we must ensure $\gamma$ is sufficiently small such that $\bar{v}$ from \eqref{eq:velocity bound} is smaller than the minimum component of $\myvar{v}_{max}$. Note that we are only concerned with $\myvar{v}_{max}$ since $\myvar{v}_{min} = - \myvar{v}_{max}$. More precisely, for $\gamma \in (0, \gamma_1^*]$, $\|\myvar{v}\|_\infty \leq \bar{v} \leq \gamma_1^* a \leq \underset{i \in \myset{N}_n}{\text{min}} v_{max_i}$, which implies that $\underset{i \in \myset{N}_n}{\text{max}} |v_i| \leq  \underset{i \in \myset{N}_n}{\text{min}} v_{max_i}$. Thus $\myvar{v} \preceq \myvar{v}_{max}$. In similar fashion, it follows that $\myvar{v}_{min} \preceq \myvar{v}$. Since this holds for all $(\myvar{q}, \myvar{v}) \in \myset{H}^\delta$, the proof is complete.
\end{pf}

\subsubsection{\textbf{Satisfaction of Input Constraints}}
\begin{figure}
\centering
\includegraphics[scale=.5]{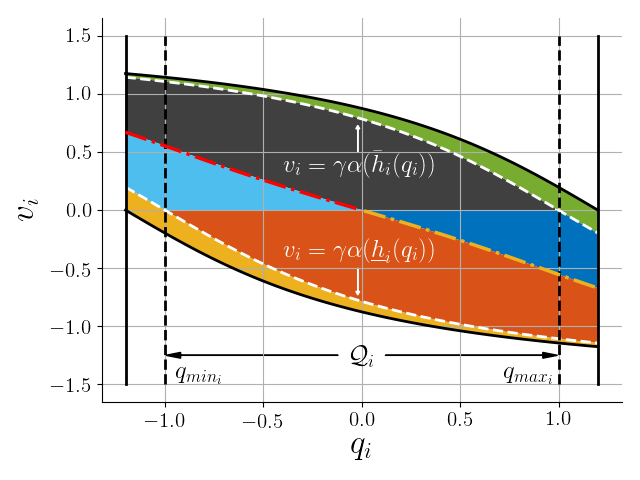}
\caption{Depiction of $\myset{H}_i$ (outlined by \rev{dashed} lines) and $\myset{H}_i^\delta$ (outlined by solid black lines). The subsets of $\myset{H}_i^\delta$ are: \RN{1} (grey), \RN{2} (blue), \RN{3} (light blue), \RN{4} (orange), \RN{5} (green), \RN{6} (yellow), \RN{7} (red dash-dotted line), and \RN{8} (yellow dash-dotted line). ZCBF parameters used in this example: $q_{max} = -q_{min} = 1.0$, $\alpha(h) = \tan^{-1}(h)$, $\gamma = 1$, $\delta = 0.2$. }\label{fig:H regions}
\end{figure}

Next, we will construct a $\tildevar{u} \in \myset{U}$ to show that there always exists a solution to \eqref{eq:Nagumos condition matrix}. However to do so, we must introduce some notation and additional terms. First we present $\rho: \myset{Q}_i^\delta \to \mathbb{R}$:
\begin{align}\label{eq:rho}
\rho(q_i) := \dfrac{\gamma}{2} \left( \alpha(\myubi{h}(q_i)) + \alpha(\mylbi{h}(q_i)) \right)
\end{align}
The function $\rho(q_i)$ defines the level set that divides $\myset{B}_i$ (see Figure \ref{fig:H regions}). More specifically, the manifold defined by: $\{ (q_i,v_i) \in \myset{H}^\delta: \myubi{b} - \rho = 0\} = \{ (q_i,v_i) \in \myset{H}^\delta: \mylbi{b} - \rho = 0\}$ is the level set for which $\myubi{b} = \mylbi{b}$. Furthermore if $\myubi{b} \leq \rho(q_i)$ then $\mylbi{b} \geq \rho(q_i)$ and vice versa. We denote the lower bound of $\rho(q_i)$ over $\myset{Q}^\delta$ as
\begin{equation}\label{eq:rho bar}
\ubar{\rho}^\delta:= \min_{i \in \myset{N}_n} \left\{ \min_{q_i \in \myset{Q}_i^\delta} \rho(q_i) \right\}
\end{equation} 
\begin{lemo}\label{lem:rho}
Suppose the conditions of Theorem \ref{thm:existenceZCBF} hold. Consider $\rho(q_i)$ and $\ubar{\rho}^\delta$ defined by \eqref{eq:rho} and \eqref{eq:rho bar}, respectively,  for a given $\gamma > 0$, $\delta \geq 0$ for  $q_i \in \myset{Q}_i^\delta$, $i \in \myset{N}_n$. Then $\rho(q_i)$ is strictly positive, and there exists a $c \in \mathbb{R}_{>0}$ such that $\ubar{\rho}^\delta \geq c$.
\end{lemo}
\begin{pf}
First, we show $\rho(q_i)$ is always strictly positive in $\myset{Q}_i \subseteq \myset{Q}_i^\delta$. From $\myubi{h} \geq 0$, $\mylbi{h} \geq 0$, and $\gamma >0$, then $\alpha(\myubi{h}(q_i))$ only equals $0$ at the boundary when $q_i = q_{max_i}$, and  $\alpha(\mylbi{h}(q_i))$ only equals $0$ at the boundary when $q_i = q_{min_i}$. Evaluation at both boundaries yields $\rho(q_{max_i}) = \rho(q_{min_i}) =  \frac{\gamma}{2}\alpha(e_i)$, for $e_i = q_{max_i}- q_{min_i}$. Since $q_{max_i} > q_{min_i}$, $e_i > 0$. Now in the interior of $\myset{Q}_i$ (i.e. $q_{min_i} < q_i < q_{max_i}$), $\alpha(\myubi{h})$ and $\alpha(\mylbi{h})$ are strictly positive. Thus there exists no such $q_i \in \myset{Q}_i$ such that $\rho(q_i) = 0$. Since $\rho$ is a continuous function on the compact set $\myset{Q}_i$, and is strictly positive, there exists some $g_i \in \mathbb{R}_{>0}$ such that $\rho(q_i) \geq g_i$ in $\myset{Q}_i$. We note that $g_i$ is independent of $\delta$.

Next, for when $\delta > 0$ and $\myset{Q}_i \subset \myset{Q}^\delta_i$, we divide $\myset{Q}_i^\delta$ into two sections: a) when $q_i \geq q_{max_i}$ and b) when $q_i \leq q_{min_i}$. For $\myset{Q}_i^\delta \setminus \myset{Q}_i$ where $q_i > q_{max_i}$, let $q_i = q_{max_i} + e$ for $e \in [0, \delta]$ such that $\rho(e) = \dfrac{\gamma}{2} \left(\alpha(-e) + \alpha(q_{max_i} - q_{min_i} - e \right)$. Then from Assumption \ref{asm:beta} it follows that $\rho(q_i) \geq d$. Finally, for $\myset{Q}_i^\delta \setminus \myset{Q}_i$ where $q_i < q_{min_i}$, let $q_i = q_{min_i} - e$ for $e \in [0, \delta]$. Similarly, it follows that $\rho(e) = \dfrac{\gamma}{2} \left(\alpha(-e) + \alpha(q_{max_i} - q_{min_i} - e \right)$, and again from Assumption \ref{asm:beta}, $\rho(q_i) \geq d$. Thus there exists some $\tilde{d}_i = \min \{ d, g_i \}$, $\tilde{d}_i \in \mathbb{R}_{>0}$ such that $\rho(q_i) \geq d_i$ on $\myset{Q}_i^\delta$. Let $c$ be the minimum of $\tilde{d}_i$ for $i \in \myset{N}_n$. By definition of $\ubar{\rho}^\delta$, it follows that $\ubar{\rho}^\delta \geq c$. 
\end{pf}
The following Lemma ensures that the sum of $\myubi{b}$ and $\mylbi{b}$ is always positive on $\myset{H}^\delta$.
\begin{lemo}\label{lem:b gap}
Suppose the conditions of Theorem \ref{thm:existenceZCBF} hold, $\gamma > 0$, $\delta \geq 0$, and consider $\rho(q_i)$ from \eqref{eq:rho}. Then $\myubi{b}(q_i,v_i) + \mylbi{b}(q_i,v_i) = 2 \rho(q_i) > 0$ for all $(q_i,v_i )\in \myset{H}_i^\delta$. Furthermore if $\myubi{b}(q_i,v_i) < \rho(q_i)$, then $\mylbi{b}(q_i,v_i) > \rho(q_i)$, and if $\mylbi{b}(q_i,v_i) <\rho(q_i)$, then $\myubi{b}(q_i,v_i) > \rho(q_i)$.
\end{lemo}
\begin{pf}
Substitution of \eqref{eq:zcbf b} into $\myubi{b} + \mylbi{b}$ yields $\myubi{b} + \mylbi{b} = 2 \rho(q_i)$. From Lemma \ref{lem:rho}, $\rho$ is strictly positive. Thus it follows that $\mylbi{b} = 2 \rho - \myubi{b} > \rho$ if $\myubi{b} < \rho$, and $\myubi{b} = 2\rho - \mylbi{b} > \rho$ if $\mylbi{b} < \rho$.
\end{pf}

Second, we introduce $\zeta^\delta_i \in \mathbb{R}$:
\begin{align}\label{eq:zeta bound}
\zeta^\delta_i := \min\{  \underset{(q_i,v_i) \in \myset{H}_i^\delta}{\text{min}}
\hspace{.03cm}  \myubi{b}, \underset{(q_i,v_i) \in \myset{H}_i^\delta}{\text{min}}
\hspace{.03cm}  \mylbi{b} \}
\end{align}
The term $\zeta_i^\delta$ is the lower bound of $\myubi{b}$ and $\mylbi{b}$ on $\myset{H}^\delta$. We denote the lower bound of $\zeta_i^\delta$ over $i \in \myset{N}_n$ as:
\begin{equation}\label{eq:zeta bar}
 \zeta^\delta = \min_{i \in \myset{N}_n} \zeta^\delta_i
\end{equation}
\begin{lemo}\label{lem:zeta}
Suppose the conditions of Theorem \ref{thm:existenceZCBF} hold, and consider $\zeta^\delta_i$, $\zeta^\delta$ defined by \eqref{eq:zeta bound} and \eqref{eq:zeta bar}, respectively, for $\gamma > 0$ and $\delta \geq 0$. Then $\zeta^\delta_i$ always exists, is non-positive, and as $\delta \searrow 0$, $\zeta^\delta \nearrow 0$.
\end{lemo}
\begin{pf}
 A solution for $\zeta_i$ always exists since $\myubi{b}$ and $\mylbi{b}$ are continuous functions over the compact set $\myset{H}_i^\delta$. Furthermore, with $\gamma > 0$, $\delta \geq 0$, there exists a coordinate $(q_{max_i} + \delta, 0) \in \myset{H}_i^\delta$ for which $\myubi{b} = -\gamma \alpha(\delta) \leq 0$. Similarly the coordinate $(q_{min_i} - \delta, 0) \in \myset{H}^\delta$ ensures $\mylbi{b} = -\gamma \alpha(\delta) \leq 0$. Since by definition \eqref{eq:zeta bound}, $\zeta_i^\delta$ is the minimum value of the minimum of $\mylbi{b}$ and $\myubi{b}$ and we have specified coordinates in $\myset{H}_i^\delta$ for which $\myubi{b}$ and $\mylbi{b}$ are non-positive, it follows that $\zeta^\delta_i$ must also be non-positive.
 
Next, from the proof of Lemma \ref{lem:velocity relations}, it follows that $-v_i \geq - \gamma \alpha(q_{max_i} - q_i + \delta)$ and $v_i \geq -\gamma \alpha(q_i - q_{min_i} + \delta)$. Thus from \eqref{eq:zcbf b}, it follows that $\myubi{b}(q_i,v_i) \geq \myubi{f}(q_i):= - \gamma \alpha(q_{max_i} - q_i + \delta) + \gamma \alpha(q_{max_i} - q_i)$ and $\mylbi{b}(q_i,v_i) \geq \mylbi{f}(q_i):=  -\gamma \alpha(q_i - q_{min_i} + \delta) + \gamma \alpha(q_i - q_{min_i})$. Thus we can re-write \eqref{eq:zeta bound} as:
\begin{align}\label{eq:zeta bound2}
\zeta^\delta_i := \min\{   & \underset{q_i \in \myset{Q}_i^\delta}{\text{min}}
\hspace{.03cm} \myubi{f}(q_i), \underset{q_i \in \myset{Q}_i^\delta}{\text{min}}
\hspace{.03cm}  \mylbi{f}(q_i) \}
\end{align}
By inspection of $\myubi{f}$ and $\mylbi{f}$, it follows that $\zeta^\delta_i = 0$ when $\delta = 0$. Furthermore, $\myubi{f}$ and $\mylbi{f}$ are non-positive, continuous, and strictly decreasing functions of $\delta$ since $\alpha$ is an extended class-$\mathcal{K}_\infty$ function and $\delta \geq 0$. Thus as $\delta \searrow 0$, $\myubi{f} \nearrow 0$ and $\mylbi{f} \nearrow 0$. Since $\zeta_i^\delta$ is the minimum of $\myubi{f}$ and $\mylbi{f}$ over $\myset{Q}_i^\delta$, it follows that as $\delta \searrow 0$, $\zeta^\delta_i \nearrow 0$. Finally, since this property holds for all $i \in \myset{N}_n$, it also holds for $\zeta^\delta$, which completes the proof.
\end{pf}
\begin{remo}
The computation of $\zeta_i^\delta$ can be done off-line as it is purely a function of the choice of $\alpha$. We explicitly define $\zeta_i^\delta$ for the following commonly used choices for $\alpha$: for $\alpha(h) = h$, $\zeta_i^\delta = -\gamma \delta$, for $\alpha(h) = \tan^{-1}(h)$, $ \zeta^\delta_i = -\gamma 2 \alpha\left(\frac{\delta}{2}\right)$, and for $\alpha(h) = h^3$, $\zeta^\delta_i = \gamma \left( \alpha\left( q_{max_i} - q_{min_i} +\delta \right) - \alpha\left(q_{max_i} - q_{min_i} + 2 \delta \right) \right)$.
\end{remo}
Finally, we divide $\myset{H}_i^\delta$ into eight regions which are outlined in Table \ref{table:H regions}, and depicted in Figure \ref{fig:H regions}. \rev{We note that Figure \ref{fig:H regions} shows the desired property that the velocity $v_i$ approaches zero as the position $q_i$ approaches the boundary of $\myset{Q}_i$ and the velocities are bounded for all $q_i \in \myset{Q}_i$.} 
\begin{table}[h!]\hspace*{-1cm}
\centering
\caption{Decomposition of $\myset{H}_i^\delta$} \label{table:H regions}
\begin{tabular}{c l}    \toprule
\RN{1} =   & $\left\{ (q_i,v_i) \in \myset{H}_i^\delta: \myubi{b}(q_i,v_i) \in [0, \rho(q_i)) \ \land\ v_i \geq 0 \right\}$  \\
\RN{2} =& $\left\{ (q_i,v_i) \in \myset{H}_i^\delta: \myubi{b}(q_i,v_i) \in [0, \rho(q_i)) \ \land \ v_i \leq 0 \right\}$ \\ 
\RN{3} =& $\left\{ (q_i,v_i) \in \myset{H}_i^\delta: \mylbi{b}(q_i,v_i) \in [0, \rho(q_i)) \ \land \ v_i \geq 0 \right\}$  \\
\RN{4} =& $\left\{ (q_i,v_i) \in \myset{H}_i^\delta: \mylbi{b}(q_i,v_i) \in [0, \rho(q_i)) \ \land \ v_i \leq 0 \right\}$   \\
\RN{5} =& $\left\{ (q_i,v_i) \in \myset{H}_i^\delta: \myubi{b}(q_i,v_i) < 0 \right\}$  \\
\RN{6} =& $\left\{ (q_i,v_i) \in \myset{H}_i^\delta: \mylbi{b}(q_i,v_i) < 0 \right\}$ \\ 
\RN{7} =& $\left\{ (q_i,v_i) \in \myset{H}_i^\delta: \mylbi{b}(q_i,v_i) = \rho(q_i) \land \ v_i \geq 0 \right\}$ \\ 
\RN{8} =& $\left\{ (q_i,v_i) \in \myset{H}_i^\delta: \mylbi{b}(q_i,v_i) = \rho(q_i) \land \ v_i \leq 0 \right\}$ \\ \bottomrule
\end{tabular}
\end{table}
We are now ready to present a candidate $\tildevar{u} : \myset{H}^\delta \to \mathbb{R}^n$ to satisfy \eqref{eq:Nagumos condition matrix} and $\tildevar{u} \in \myset{U}$:
\begin{align}\label{eq:u exists}
\tildevar{u}(\myvar{q}, \myvar{v}) :=  \rev{G^+(\myvar{q})} \left( \myvar{\mu}(\myvar{q}, \myvar{v}) + \myvar{\chi}(\myvar{q}, \myvar{v}) + \myvar{\psi}(\myvar{q}, \myvar{v}) \right) \notag \\
\rev{- \myvar{f}_1(\myvar{q}, \myvar{v})  - \myvar{f}_2(\myvar{q},\myvar{v}) - \myvar{f}_3(\myvar{q})} &
\end{align}
where
\begin{align}\label{eq:u exists mu}
\mu_i(q_i,v_i) := \begin{cases}
- \gamma \dfrac{\partial \alpha}{\partial \myubi{h}}(q_i) v_i, & \text{ if } (q_i,v_i) \in \RN{1} \cup \RN{5} \cup \RN{7}  \\
0, & \text{ if } (q_i,v_i) \in \RN{2} \cup \RN{3} \\
- \gamma \dfrac{\partial \alpha}{\partial \mylbi{h}}(q_i) v_i,& \text{ if } (q_i,v_i) \in \RN{4} \cup \RN{6} \cup \RN{8} \\
 \end{cases}
\end{align}
\begin{align}\label{eq:u exists chi}
\chi_i(q_i,v_i) :=
\begin{cases}
0, \text{ if } (q_i,v_i) \in \RN{1} \cup \RN{2} \cup \RN{3} \cup \RN{4} \cup \RN{7} \cup \RN{8}  \\
\nu \beta (\myubi{b}(q_i,v_i)), \text{ if } (q_i,v_i) \in \RN{5}  \\
-\nu \beta (\mylbi{b}(q_i,v_i)), \text{ if } (q_i,v_i) \in \RN{6} 
\end{cases}
\end{align}
\begin{align}\label{eq:u exists psi}
\psi_i(q_i,v_i) :=
\begin{cases}
-\bar{\eta}, &\text{ if } (q_i,v_i) \in \RN{1} \cup \RN{2} \cup \RN{5} \\
\bar{\eta}, &\text{ if }  (q_i,v_i) \in \RN{3} \cup \RN{4} \cup \RN{6} \\
0, &\text{ if } (q_i,v_i) \in \RN{7} \cup \RN{8}
\end{cases},
\end{align} 
$\myvar{\chi}(\myvar{q}, \myvar{v})$ $= [\chi_1(q_1,v_1)$ $, ...,$ $\chi_n(q_n,v_n)]^T$,  $\myvar{\mu}$ $:= [\mu_1(q_1,v_1)$ $,...,$ $\mu_n(q_n,v_n)]^T$, and $\myvar{\psi}$ $:= [\psi_1(q_1,v_1)$ $,...,$ $\psi_n(q_n,v_n)]^T$. We note that $\tildevar{u}$ is well-defined over all of $\myset{H}^\delta$. Furthermore, $\tildevar{u}$ is discontinuous over $\myset{H}^\delta$. We address discontinuities in a sampled-data fashion as will be discussed later.

Our first task is to ensure that $\tildevar{u} \in \myset{U}$ for all $(\myvar{q}, \myvar{v}) \in \myset{H}^\delta$. We do this by bounding $\gamma$ using:
\begin{equation}\label{eq:gamma star}
\gamma_2^* = \underset{\begin{subarray}{c} \myvar{q} \in \myset{Q}^\delta \\ \rev{j \in \myset{N}_m} \end{subarray}}{\text{min}} \dfrac{-d_{\rev{j}}(\myvar{q}) + \sqrt{ d_{\rev{j}}^2 - 4 c_{\rev{j}}(\myvar{q}) }}{2}
\end{equation}
where
\begin{equation}
d_{\rev{j}}(\myvar{q}) = \dfrac{f_{\rev{j}}}{\|\myvar{e}_{\rev{j}}^T G^+(\myvar{q})\|_\infty y(\myvar{q})  + k_c a}
\end{equation} 
\begin{equation}
c_{\rev{j}}(\myvar{q}) = \dfrac{| \rev{\myvar{e}_{\rev{j}}^T\myvar{f}_3(\myvar{q})}| + (\varepsilon + \bar{\eta})\| \myvar{e}_{\rev{j}}^T \rev{G(\myvar{q})^+ }\|_\infty - u_{max_{\rev{j}}}}{\|\myvar{e}_{\rev{j}}^T \rev{G(\myvar{q})^+}\|_\infty y(\myvar{q})a  + k_c a^2},
\end{equation}
$y(\myvar{q}) = \max_{i \in \myset{N}_n} \left\{ \dfrac{\partial \alpha}{\partial \myubi{h}}(q_i), \dfrac{\partial \alpha}{\partial \mylbi{h}}(q_i) \right\} $, and $f_{\rev{j}} \in \mathbb{R}$ is \rev{from Property \ref{prop:F}}.
The idea behind $\gamma_2^*$ is that as $\gamma$ decreases, the system velocity will decrease and ensure the system inertia is not too large to exceed the limitations of the system's actuators.

Similarly, we define the upper bound $\nu_2^*$ to ensure $\|\myvar{\chi}\|_\infty \leq \varepsilon$ to respect actuator constraints in $\myset{H}^\delta \setminus \myset{H}$:
\begin{equation}\label{eq:nu star 2}
\nu_2^* := \dfrac{\varepsilon}{|\beta(\zeta^\delta)| }
\end{equation}
In the event that $\delta = 0$, then clearly $\nu_2^* = \infty$, which implies that the choice of $\nu$ is not upper bounded.

Satisfaction of $\tildevar{u} \in \myset{U}$ is formally stated in the following Lemma:
\begin{lemo}\label{lem:actuator constraints}
Suppose the conditions of Theorem \ref{thm:existenceZCBF} hold. Consider $\tildevar{u}: \myset{H}^\delta \to \mathbb{R}^n$ defined by \eqref{eq:u exists}, $\gamma_2^*$ defined by \eqref{eq:gamma star} and $\nu_2^*$ defined by \eqref{eq:nu star 2} with $\varepsilon$ from Assumption \ref{asm:control}. Then $\gamma_2^*$ always exists and is strictly positive, and $\nu_2^*$ is always strictly positive and bounded if $\delta > 0$, otherwise $\nu_2^* = +\infty$ if $\delta = 0$. Furthermore, if $\gamma \in (0, \gamma_2^*]$, $\nu \in (0, \nu_2^*]$ for $\delta > 0$ otherwise $\nu > 0$ if $\delta = 0$, then $\tildevar{u} \in \myset{U}$ for all $(\myvar{q}, \myvar{v}) \in \myset{H}^\delta$.
\end{lemo}
\begin{pf}
We start with ensuring existence of strictly positive $\gamma_2^*$ and $\nu_2^*$. Existence and positivity of $\nu_2^*$ follows trivially from \eqref{eq:nu star 2} and Assumption \ref{asm:control} for $\delta > 0$. If $\delta = 0$, then $\nu_2^* = + \infty$  follows trivially from \eqref{eq:nu star 2}. Since we chose $\varepsilon$ from Assumption \ref{asm:control}, it follows that $c_{\rev{j}}(\myvar{q}) < 0$ in \eqref{eq:gamma star}, and so $\gamma_2^*$ is real and positive.

Now we ensure the satisfaction of the actuator constraints $\tildevar{u} \in \myset{U}$. Since $\myvar{u}_{max} = -\myvar{u}_{min}$, we write the actuator constraint condition as $|u_{\rev{j}}| - u_{max_{\rev{j}}} \leq 0$ for all $\rev{j} \in \rev{\myset{N}_m}$. Substitution of $\tildevar{u}$ into $|u_{\rev{j}}| - u_{max_{\rev{j}}} \leq 0$ yields:
\begin{align*}
 | \myvar{e}_{\rev{j}}^T \left( \rev{G}^+ \left( \myvar{\mu}(\myvar{q}, \myvar{v}) + \myvar{\chi}(\myvar{q}, \myvar{v}) + \myvar{\psi}(\myvar{q}, \myvar{v})\right) \rev{- \myvar{f}_1(\myvar{q}, \myvar{v}) }\right. \\
\left. - \rev{ \myvar{f}_2(\myvar{q},\myvar{v}) - \myvar{f}_3(\myvar{q})} \right)  | - u_{max_{\rev{j}}} \leq 0 
\end{align*}

First we consider the case $\delta > 0$ such that $\nu_2^* < \infty$. By choice of $\nu \in (0, \nu_2^*]$, $\nu | \beta(\zeta^\delta)| \leq \varepsilon$.  It straightforward to see that the lower bound on $\myubi{b}$ is reached in $\RN{5}$ when $\myubi{b} < 0$, and similarly $\mylbi{b}$ reaches its lower bound in $\RN{6}$ when $\mylbi{b}< 0$, for $i \in \myset{N}_n$. From \eqref{eq:zeta bound} and \eqref{eq:zeta bar} it follows that $|\beta(\myubi{b})| \leq |\beta(\zeta^\delta)|$, $|\beta(\mylbi{b}) |\leq |\beta(\zeta^\delta)|$ in \RN{5} and \RN{6}, respectively. From \eqref{eq:u exists chi}, in \RN{1}-\RN{4}, \RN{7}, and \RN{8}, $\chi_i = 0$. In \RN{5}, $|\chi_i| \leq \nu |\beta(\myubi{b})| \leq |\beta(\zeta^\delta)| \leq \varepsilon$. In \RN{6}, $|\chi_i| \leq \nu |\beta(\mylbi{b})| \leq |\beta(\zeta^\delta)| \leq \varepsilon$. Thus $\|\myvar{\chi}\|_\infty \leq \varepsilon$ on $\myset{H}^\delta$. It is also straightforward to see that $\|\myvar{\psi}\|_\infty \leq \bar{\eta}$.

From Properties \ref{prop:C}, \rev{\ref{prop:F},} and Lemma \ref{lem:velocity relations}, it follows that for all $(\myvar{q}, \myvar{v}) \in \myset{H}^\delta$, $\|\rev{\myvar{f}_1(\myvar{q}, \myvar{v})\|_\infty \leq k_c \|\myvar{v}\|_\infty^2} \leq k_c  \bar{v}^2 = k_c  \gamma^2 a^2$ \rev{and $\|\myvar{e}_j^T \myvar{f}_2(\myvar{q}, \myvar{v}) \| \leq f_j \|\myvar{v}\|_\infty \leq f_j \gamma a$, for $j \in \myset{N}_m$}. By definition of $y(\myvar{q})$, it follows that $\| \myvar{\mu} \|_\infty \leq \gamma^2 y(\myvar{q}) a$. Substitution of \rev{$\|\myvar{f}_1(\myvar{q}, \myvar{v})\|_\infty \leq k_c \gamma^2 a^2$}, $\| \myvar{\chi}\|_\infty \leq \varepsilon$, $\|\myvar{\psi}\|_\infty \leq \bar{\eta}$, $\| \myvar{\mu} \|_\infty \leq \gamma^2 y(\myvar{q}) a$, \rev{$\| \myvar{e}_{\rev{j}}^T \myvar{f}_2(\myvar{q}, \myvar{v})\| \leq f_{\rev{j}}  \gamma a$}, and application of the triangle inequality yields the following sufficient condition for guaranteeing that $\tildevar{u} \in \myset{U}$:
\begin{align*}
\gamma^2 \left( \|\myvar{e}_{\rev{j}}^T \rev{G(\myvar{q})^+} \|_\infty y(\myvar{q})a + k_c a^2 \right)  + \gamma f_{\rev{j}} a + | \rev{\myvar{e}_{\rev{j}}^T \myvar{f}_3(\myvar{q})}| \\
 + (\varepsilon + \bar{\eta})\| \myvar{e}_{\rev{j}}^T \rev{G(\myvar{q})^+} \|_\infty 
- u_{max_{\rev{j}}} \leq 0 &
\end{align*}
Application of the standard quadratic formula to solve for $\gamma$ (at equality) for all $\rev{j} \in \rev{\myset{N}_{m}}$ yields \eqref{eq:gamma star}. Thus if $\gamma = \gamma_2^*$, then $\tildevar{u} \in \myset{U}$. Furthermore, it is trivial to see that any $\gamma \in (0, \gamma_2^*]$ also ensures $\tildevar{u} \in \myset{U}$. 
In the event that $\delta = 0$, then the sets \RN{5} and \RN{6} are in fact empty. Thus $\myvar{\chi} = 0$ on $\myset{H}^\delta$, which satisfies $\| \myvar{\chi} \|_\infty \leq \varepsilon$ and the previous analysis ensures that if $\gamma \in (0, \gamma_2^*]$, $\nu > 0$, then $\tildevar{u} \in \myset{U}$.
\end{pf}

\subsubsection{\textbf{Non-Conflicting ZCBFs}}
Next, we design $\gamma_3^*$, $\delta^*$, $\nu_1^*$, and $\eta^*$ to ensure non-conflicting ZCBF conditions. The candidate ZCBFs require the $2n$ conditions from \eqref{eq:Nagumo rel 2 condition} to be satisfied at all times on $\myset{H}^\delta$. We substitute \eqref{eq:u exists} into \eqref{eq:Nagumo rel 2 condition}, which yields:
\begin{align}
&\mu_i + \chi_i + \psi_i + \gamma \frac{\partial \alpha}{\partial \myubi{h}} v_i - \nu \beta(\myubi{b}) + \bar{\eta} \leq 0 \label{eq:Nagumo condition for control1} \\
&\mu_i+ \chi_i + \psi_i+ \gamma \frac{\partial \alpha}{\partial \mylbi{h}} v_i  +  \nu \beta(\mylbi{b}) - \bar{\eta} \geq 0 \label{eq:Nagumo condition for control2}
\end{align}
for $i \in \myset{N}_n$. Thus satisfaction of \eqref{eq:Nagumo condition for control1} and \eqref{eq:Nagumo condition for control2} over all $i \in \myset{N}_n$ ensures \eqref{eq:Nagumo rel 2 condition} holds. We must now ensure there are no conflicting conditions such that $\tildevar{u}$ can satisfy \eqref{eq:Nagumo condition for control1} and \eqref{eq:Nagumo condition for control2} simultaneously for all $i \in \myset{N}_n$. 

We now define the following upper bound $\gamma_3^*$ to prevent conflict in \eqref{eq:Nagumos condition matrix}:
\begin{align}\label{eq:gamma star 2}
\gamma_3^* :=  \sqrt{ \dfrac{\varepsilon}{L a}}
\end{align}
where $a$ is defined in \eqref{eq:velocity bound} and $L \in \mathbb{R}_{>0}$ is the Lipschitz constant of $\alpha$ for all $\myubi{h}(q_i)$, $\mylbi{h}(q_i)$  for all $\myvar{q} \in \myset{Q}^\delta$.   

Next we design the lower bound $\nu_1^*$ to ensure there always exists a control in $\myset{H}^\delta \setminus \myset{H}$ to satisfy the ZCBF conditions:
\begin{equation}\label{eq:nu star 1}
\nu_1^* :=  \dfrac{\gamma^2 L a}{\beta(\ubar{\rho}^\delta)}
\end{equation}

In the following Lemma we show that for a sufficiently small $\delta$ and choice of $\gamma \in (0, \gamma_3^*]$, the previous designs of $\nu_1^*$, $\nu_2^*$ are well-defined such that $\nu_1^* < \nu_2^*$:
\begin{lemo}\label{lem:delta}
Suppose the conditions of Theorem \ref{thm:existenceZCBF} hold and consider $\gamma_3^*$, $\nu_1^*$, $\nu_2^*$ defined, respectively, by \eqref{eq:gamma star 2}, \eqref{eq:nu star 1}, \eqref{eq:nu star 2}, for $\delta \geq 0$. Then $\gamma_3^*$ always exists and is strictly positive. Furthermore, for $\gamma \in (0, \gamma_3^*]$, there exists a $\delta^* \in \mathbb{R}_{>0}$ that satisfies the following conditions:
\begin{align}\label{eq:delta_star}
|\beta(\zeta^\delta)| < \beta(\ubar{\rho}^\delta), \forall \delta \in [0, \delta^*]
\end{align}
and for $\delta \in [0, \delta^*]$,  $\nu_1^* > 0$, $\nu_2^* > 0$, and $\nu_1^* < \nu_2^*$. 
\end{lemo}
\begin{pf}
First, we ensure $\gamma_3^*$ is strictly positive. Since $\alpha$ is continuously differentiable there always exists a Lipschitz constant $L >0$ and with $a >0$ it is straightforward to see that $\dfrac{\varepsilon}{L a}$, and thus $\gamma_3^*$, is strictly positive.

Existence of \eqref{eq:delta_star} follows from Lemmas \ref{lem:zeta} and \ref{lem:rho} and the fact that $\beta$ is an extended class-$\mathcal{K}_\infty$ function such that as $\delta \searrow 0$, $|\beta(\zeta^\delta)| \searrow 0$. Furthermore, since $\ubar{\rho}^\delta \geq c$ from Lemma \ref{lem:rho}, there exists a sufficiently small $\delta' \in \mathbb{R}_{>0}$ such that $| \beta(\zeta^\delta) | < \beta(c) \leq \beta(\ubar{\rho}^\delta)$. Let $\delta^* = \delta'$. Since $\ubar{\rho}^\delta$ is lower bounded by $c$ and $\zeta^\delta$ will continue to approach $0$, it follows that the choice of $\delta^*$ satisfies \eqref{eq:delta_star}.  

Next, we show $\nu_1^*$ is well-defined such that $\nu_1^* < \nu_2^*$. Since $\rho(q_i)$ (and thus $\ubar{\rho}^\delta$) is strictly positive from Lemma \ref{lem:rho}, $\nu_1^*$ is strictly positive. For $\gamma \in (0, \gamma_3^*]$, it follows that $\nu_1^* = \dfrac{\gamma^2 L a}{\beta(\ubar{\rho}^\delta)} \leq \dfrac{\varepsilon}{\beta(\ubar{\rho}^\delta)}$. Now for $\delta \in [0, \delta^*]$, it follows that $|\beta(\zeta^\delta)| <\beta(\ubar{\rho}^\delta)$ such that $\nu_1^* \leq \dfrac{\varepsilon}{\beta(\ubar{\rho}^\delta)} < \dfrac{\varepsilon}{|\beta(\zeta^\delta)|} :=\nu_2^*$.
\end{pf}

The final component to the proper design of $\gamma$ and $\nu$ is the design of $\bar{\eta}$. Recall that $\bar{\eta}$ is an added robustness margin to handle sampling time effects\footnote{\rev{This robustness margin can also address disturbances on the system dynamics, see \cite{Lindemann2019}.}}. In this respect, $\bar{\eta}$ must be sufficiently small (i.e., the sampling frequency must be sufficiently fast) such that no conflict occurs when attempting to simultaneously satisfy \eqref{eq:Nagumo condition for control1} and \eqref{eq:Nagumo condition for control2}. We define the upper bound on $\bar{\eta}$ as:
\begin{equation}\label{eq:eta star}
\eta^* := \dfrac{\nu \beta(\ubar{\rho}^\delta) - \gamma^2 L a}{2}
\end{equation}
\begin{lemo}\label{lem:eta star}
Suppose the conditions of Theorem \ref{thm:existenceZCBF} hold and consider $\gamma_3^*$, $\nu_1^*$, $\nu_2^*$ defined, respectively, by \eqref{eq:gamma star 2}, \eqref{eq:nu star 1}, \eqref{eq:nu star 2}, for $\gamma > 0$, $\delta \geq 0$. If $\delta \in [0, \delta^*]$, $\gamma \in (0,  \gamma_3^*]$, $\nu \in [\nu_1^*, \nu_2^*]$ for $\delta > 0$ otherwise $\nu \geq \nu_1^*$ if $\delta = 0$, then $\eta^*$ is non-negative. Furthermore, if $\nu > \nu_1^*$ then $\eta^*$ is strictly positive.
\end{lemo}
\begin{pf}
By Lemma \ref{lem:delta}, it follows that $\nu_1^* < \nu_2^*$. For $\nu \geq \nu_1^*$, then $\nu \geq \dfrac{\gamma^2 L a}{\beta(\ubar{\rho}^\delta)}$ and it follows that $\nu \beta(\ubar{\rho}^\delta) - \gamma^2 L a \geq 0$. Thus $\eta^*$ from \eqref{eq:eta star} must be non-negative. Similarly if $\nu > \nu_1^*$ then $\nu > \dfrac{\gamma^2 L a}{\beta(\ubar{\rho}^\delta)}$ and so $\nu \beta(\ubar{\rho}^\delta) - \gamma^2 L a > 0$, and so $\eta^*$ is strictly positive.
\end{pf}
The following Lemma shows that the choice of $\gamma \in (0, \gamma_3^*]$, $\nu \in [\nu_1^*, \nu_2^*]$, and $\bar{\eta} \in [0, \eta^*]$ prevents conflict between the ZCBF conditions:
\begin{lemo}\label{lem:nonconflicting}
Suppose the conditions of Theorem \ref{thm:existenceZCBF} hold and consider $\gamma_3^*$, $\nu_1^*$, $\nu_2^*$, $\delta^*$, and $\eta^*$ defined, respectively, by  \eqref{eq:gamma star 2}, \eqref{eq:nu star 1}, \eqref{eq:nu star 2}, \eqref{eq:delta_star}, \eqref{eq:eta star}. For $ \delta \in [0, \delta^*]$  $\gamma \in (0,  \gamma_3^*]$, $\nu \in [\nu_1^*, \nu_2^*]$ for $\delta > 0$ otherwise $\nu \geq \nu_1^*$ if $\delta = 0$, and $\bar{\eta} \in [0, \eta^*]$, then the following conditions are always satisfied:
\begin{align}\label{eq:nonconflicting condition1}
- \gamma \left( \dfrac{\partial \alpha}{\partial \myubi{h}}(q_i) - \dfrac{\partial \alpha}{\partial \mylbi{h}}(q_i) \right) v_i  - 2 \bar{\eta} + \nu \beta(\ubar{\rho}^\delta) > 0, \notag \\ \ \forall (q_i,v_i)\in \myset{H}_i^\delta
\end{align}
\begin{align}\label{eq:nonconflicting condition2}
 \gamma \dfrac{\partial \alpha}{\partial \mylbi{h}}(q_i)  v_i - 2 \bar{\eta} + \nu \beta(\ubar{\rho}^\delta) \geq 0, \ \forall (q_i,v_i)\in \RN{2}
\end{align}
\begin{align}\label{eq:nonconflicting condition3}
 \gamma \dfrac{\partial \alpha}{\partial \myubi{h}}(q_i)  v_i + 2 \bar{\eta} - \nu \beta(\ubar{\rho}^\delta) \leq 0 \ \forall (q_i,v_i)\in \RN{3}
\end{align}
\end{lemo}
\begin{pf}
To show satisfaction \eqref{eq:nonconflicting condition1}, we note the following bounds for $(q_i,v_i) \in \myset{H}_i^\delta$: 
\begin{align*}
-\gamma \left( \dfrac{\partial \alpha}{\partial \myubi{h}}(q_i) - \dfrac{\partial \alpha}{\partial \mylbi{h}}(q_i) \right)v_i &\geq - \gamma \ | \dfrac{\partial \alpha}{\partial \myubi{h}}(q_i) - \dfrac{\partial \alpha}{\partial \mylbi{h}}(q_i)| \ |v_i|  \\
& > - \gamma \min\{  \dfrac{\partial \alpha}{\partial \myubi{h}}(q_i), \dfrac{\partial \alpha}{\partial \mylbi{h}}(q_i) \} \bar{v} \\
& \geq  - \gamma^2 L a
\end{align*}
where $| \dfrac{\partial \alpha}{\partial \myubi{h}}(q_i) - \dfrac{\partial \alpha}{\partial \mylbi{h}}(q_i)| < \min \{  \dfrac{\partial \alpha}{\partial \myubi{h}}(q_i), \dfrac{\partial \alpha}{\partial \mylbi{h}}(q_i) \}$ holds because $\alpha$ is strictly increasing. Also, the bound: $|v_i| \leq \bar{v} = \gamma a$ follows from Lemma \ref{lem:velocity relations}.
From Lemmas \ref{lem:delta} and \ref{lem:eta star}, the choices for $\nu \in [ \nu_1^*, \nu_2^*]$ for $\delta > 0$ otherwise $\nu \geq \nu_1^*$ if $\delta = 0$, $\eta \in [0, \eta^*]$ are well-defined. Satisfaction of \eqref{eq:nonconflicting condition1} follows by substution of \eqref{eq:eta star} with the above bound.

Next we show satisfaction of \eqref{eq:nonconflicting condition2}. Using the aforementioned bounds (for $v_i \leq 0$ in \RN{2}) yields: $ \gamma \dfrac{\partial \alpha}{\partial \mylbi{h}}(q_i)  v_i \geq - \gamma^2 L a $. Thus substitution of \eqref{eq:eta star} along with the previous bound ensures \eqref{eq:nonconflicting condition2} is satisfied.

Satisfaction of \eqref{eq:nonconflicting condition3} is similar to the above cases. For $v_i \geq 0$ in \RN{3}, it follows that  $ \gamma \dfrac{\partial \alpha}{\partial \myubi{h}}(q_i)  v_i \leq  \gamma^2 L a$. Thus \eqref{eq:nonconflicting condition3} is satisfied with this bound and appropriate substitution of \eqref{eq:eta star}.
\end{pf}
Note that the requirements of Lemma \ref{lem:nonconflicting} are the main components to avoid conflict such that \eqref{eq:nonconflicting condition1} and \eqref{eq:nonconflicting condition2} always hold simultaneously. The formal guarantees of non-conflicting conditions are found in the following proof of Theorem \ref{thm:existenceZCBF}.

We are now ready to present the proof of Theorem \ref{thm:existenceZCBF}:
\begin{pf}[Proof of Theorem \ref{thm:existenceZCBF}]
We must show that there exists a $\myvar{u} \in \myset{U}$ such that \eqref{eq:Nagumos condition matrix} holds for all $(\myvar{q}, \myvar{v})$ in $\myset{H}^\delta$. The proof is composed of four parts. First, we ensure the existence of $\gamma_1^*$, $\gamma_2^*$, $\gamma_3^*$, $\nu_1^*$, $\nu_2^*$, and define the upper bounds on $\delta$ and $\bar{\eta}$. Second, we show that a candidate $\tildevar{u} \in \myset{U}$ is well-defined in $\myset{H}^\delta$. Third, we ensure that $\myvar{v} \in \myset{V}$. Fourth, we show that $\tildevar{u}$ satisfies \eqref{eq:Nagumos condition matrix} on $\myset{H}^\delta$. 

1) Let $\gamma_1^*$, $\gamma_2^*$, $\gamma_3^*$ be defined by \eqref{eq:gamma star 3}, \eqref{eq:gamma star}, and \eqref{eq:gamma star 2}, respectively. For $\delta$, $\bar{\eta} \geq 0$ satisfying Assumption \ref{asm:control}, it follows that $\gamma_1^*$ exists and is strictly positive from \eqref{eq:gamma star 3}. Lemmas \ref{lem:actuator constraints} and \ref{lem:delta} ensure $\gamma_2^*$ and $\gamma_3^*$ always exists and are strictly positive. Lemma \ref{lem:actuator constraints} also ensures $\nu_2^*$ exists and is strictly positive for $\delta > 0$, and otherwise $\nu_2^* = +\infty$ if $\delta = 0$. For $\gamma \in (0, \min \{ \gamma_1^*, \gamma_2^*, \gamma_3^*\}]$, Lemma \ref{lem:delta} ensures that $\delta^*$ is well-defined and strictly positive. We restrict $\delta$ such that $\delta \in [0, \delta^*]$. Now Lemma \ref{lem:delta} ensures $\nu_1^*$ is strictly positive and $\nu_1^* < \nu_2^*$. Finally, Lemma \ref{lem:eta star} ensures that for $\nu \in [\nu_1^*, \nu_2^*]$ if $\delta > 0$ otherwise $\nu \geq \nu_1^*$ if $\delta = 0$, $\eta^*$ is non-negative. We restrict $\bar{\eta}$ such that $\bar{\eta} \in [0, \eta^*]$.

2) Let $\tildevar{u}$ from \eqref{eq:u exists} be the candidate control law. From Lemma \ref{lem:actuator constraints}, it follows that $\tildevar{u} \in \myset{U}$ for all $(\myvar{q}, \myvar{v}) \in \myset{H}^\delta$. 

3) By Lemma \ref{lem:velocity consat}, it follows that for any $(\myvar{q}, \myvar{v}) \in \myset{H}^\delta$, $\myvar{v} \in \myset{V}$.

4) Here we ensure that $\tildevar{u}$ satisfies \eqref{eq:Nagumos condition matrix}. Substitution of \eqref{eq:u exists} into \eqref{eq:Nagumos condition matrix} yields \eqref{eq:Nagumo condition for control1} and \eqref{eq:Nagumo condition for control2} for $i \in \myset{N}_n$. Now we investigate \eqref{eq:Nagumo condition for control1} and \eqref{eq:Nagumo condition for control2} over $\myset{H}^\delta$ by decomposing $\myset{H}^\delta$ into the eight regions from Table \ref{table:H regions} and substituting $\mu_i$, $\chi_i$, and $\psi_i$ appropriately:

\textbf{\RN{1}}: $ \left[ \mu_i = -\gamma \dfrac{\partial \alpha}{\partial \myubi{h}}(q_i), \ \chi_i = 0, \ \psi_i = -\bar{\eta}\right]$. The left-hand-side of \eqref{eq:Nagumo condition for control1} yields: $- \nu \beta(\myubi{b})$ which is non-positive in \RN{1}. The left-hand-side of \eqref{eq:Nagumo condition for control2} yields:
\begin{align*}
- \gamma \left( \dfrac{\partial \alpha}{\partial \myubi{h}}(q_i) - \dfrac{\partial \alpha}{\partial \mylbi{h}}(q_i) \right) v_i -2 \bar{\eta} + \nu \beta(\mylbi{b}(q_i,v_i))
\end{align*}
For $\myubi{b} < \rho$ , it follows that $\mylbi{b} > \rho \geq \bar{\rho}^\delta$ from Lemma \ref{lem:b gap} and \eqref{eq:rho bar}. Thus $\nu \beta(\mylbi{b}) > \nu \beta(\rho) \geq \nu \beta(\ubar{\rho}^\delta)$ since $\beta$ is an extended class-$\mathcal{K}_\infty$ function. Substitution of $\nu \beta(\mylbi{b}) > \nu \beta(\ubar{\rho}^\delta)$ into the above inequality is strictly greater than the left-hand-side of \eqref{eq:nonconflicting condition1}, which by Lemma \ref{lem:nonconflicting} ensures \eqref{eq:Nagumo condition for control2} holds. Thus \eqref{eq:Nagumo condition for control1} and \eqref{eq:Nagumo condition for control2} hold in \RN{1}.

\textbf{\RN{2}}:  $ \left[ \mu_i = 0, \ \chi_i = 0, \ \psi_i = -\bar{\eta}\right]$. The left-hand-side of \eqref{eq:Nagumo condition for control1} yields: $\gamma \dfrac{\partial \alpha}{\partial \myubi{h}}(q_i) v_i  - \nu \beta(\myubi{b})$, for which $ \gamma \dfrac{\partial \alpha}{\partial \myubi{h}}(q_i) v_i$ is non-positive, since $\alpha$ is strictly increasing and $v_i \leq 0$, such that \eqref{eq:Nagumo condition for control1} holds. The left-hand-side of \eqref{eq:Nagumo condition for control2} is strictly greater than the left-hand-side of \eqref{eq:nonconflicting condition2} since $\mylbi{b} > \rho \geq \ubar{\rho}^\delta$ in \RN{2} from Lemma \ref{lem:b gap} and so $\nu \beta(\mylbi{b}) >  \nu \beta(\ubar{\rho}^\delta)$. Thus by Lemma \ref{lem:nonconflicting}, \eqref{eq:Nagumo condition for control2} holds.

\textbf{\RN{3}}:  $ \left[ \mu_i = 0, \ \chi_i = 0, \ \psi_i = \bar{\eta}\right]$. The left-hand-side of \eqref{eq:Nagumo condition for control1} is strictly less than the left-hand-side of \eqref{eq:nonconflicting condition3} since $\myubi{b} > \rho$ in \RN{3} by Lemma \ref{lem:b gap} and so $- \nu \beta(\mylbi{b}) < -\nu \beta(\rho) \leq - \nu \beta(\ubar{\rho}^\delta)$. Thus by Lemma \ref{lem:nonconflicting}, \eqref{eq:Nagumo condition for control1} holds. The left-hand-side of \eqref{eq:Nagumo condition for control2} yields: $\gamma \dfrac{\partial \alpha}{\partial \mylbi{h}}(q_i) v_i + \nu \beta(\mylbi{b})$,
for which $ \gamma \dfrac{\partial \alpha}{\partial \mylbi{h}}(q_i) v_i$ is non-negative, since $\alpha$ is strictly increasing and $v_i \geq 0$, and $\mylbi{b} \geq 0$ by definition of \RN{3} such that \eqref{eq:Nagumo condition for control2} holds.

\textbf{\RN{4}}:  $ \left[ \mu_i = -\gamma \dfrac{\partial \alpha}{\partial \mylbi{h}}(q_i), \ \chi_i = 0, \ \psi_i = \bar{\eta}\right]$. The left-hand-side \eqref{eq:Nagumo condition for control1} yields:
\begin{align*}
- \gamma \left( \dfrac{\partial \alpha}{\partial \mylbi{h}}(q_i) - \dfrac{\partial \alpha}{\partial \myubi{h}}(q_i) \right) v_i + 2\bar{\eta} - \nu \beta(\myubi{b})
\end{align*}
Since $\myubi{b} > \rho(q_i)$ in \RN{4} from Lemma \ref{lem:b gap}, it follows that $- \nu \beta(\myubi{b}) \leq - \nu \beta(\ubar{\rho}^\delta)$ such that substitution in the above inequality and Lemma \ref{lem:nonconflicting} ensures the above inequality is non-positive and so \eqref{eq:Nagumo condition for control1} holds. The left-hand-side of \eqref{eq:Nagumo condition for control2} yields $\nu \beta(\mylbi{b})$, which is non-negative in \RN{4}, and so \eqref{eq:Nagumo condition for control2} holds.

\textbf{\RN{5}}:  $ \left[ \mu_i = -\gamma \dfrac{\partial \alpha}{\partial \myubi{h}}(q_i), \ \chi_i = \nu \beta(\myubi{b}), \ \psi_i = -\bar{\eta}\right]$. The left-hand-side of \eqref{eq:Nagumo condition for control1} equals $0$ and thus \eqref{eq:Nagumo condition for control1} is satisfied. The left-hand-side of \eqref{eq:Nagumo condition for control2} yields:
\begin{align*}
- \gamma \left( \dfrac{\partial \alpha}{\partial \myubi{h}}(q_i) - \dfrac{\partial \alpha}{\partial \mylbi{h}} \right) v_i - 2 \bar{\eta} + \nu \beta(\myubi{b}) + \nu \beta(\mylbi{b}) \\
\geq - \gamma \left( \dfrac{\partial \alpha}{\partial \myubi{h}}(q_i) - \dfrac{\partial \alpha}{\partial \mylbi{h}} \right) v_i -2 \bar{\eta} + \nu \beta(\rho(q_i))
\end{align*}
We note that the above inequality holds due to Assumption \ref{asm:beta} since $\myubi{b} + \mylbi{b} = 2 \rho(q_i)$ (via Lemma \ref{lem:b gap}), $\myubi{b} < 0$ in \RN{5}, and thus $\beta(\myubi{b}) + \beta(\mylbi{b}) \geq \beta(\rho(q_i))$. Since $\rho \geq \ubar{\rho}^\delta$, \eqref{eq:Nagumo condition for control2} is satisfied from Lemma \ref{lem:nonconflicting}.

\textbf{\RN{6}}: $ \left[ \mu_i = -\gamma \dfrac{\partial \alpha}{\partial \mylbi{h}}(q_i), \ \chi_i = -\nu \beta(\mylbi{b}), \ \psi_i = \bar{\eta}\right]$. The left-hand-side of \eqref{eq:Nagumo condition for control1} yields:
\begin{align*}
- \gamma \left( \dfrac{\partial \alpha}{\partial \mylbi{h}}(q_i) - \dfrac{\partial \alpha}{\partial \myubi{h}} \right) v_i + 2 \bar{\eta} - \nu \beta(\myubi{b}) - \nu \beta(\mylbi{b}) \\
\leq - \gamma \left( \dfrac{\partial \alpha}{\partial \mylbi{h}}(q_i) - \dfrac{\partial \alpha}{\partial \myubi{h}} \right) v_i + 2 \bar{\eta} - \nu \beta(\rho(q_i))
\end{align*}
Again, the above inequality holds due to Lemma \ref{lem:b gap} and Assumption \ref{asm:beta} such that $-\beta(\myubi{b}) -\beta(\mylbi{b}) \leq -\beta(\rho(q_i))$. Thus \eqref{eq:Nagumo condition for control1} holds from Lemma \ref{lem:nonconflicting}. The left-hand-side of \eqref{eq:Nagumo condition for control2} equals $0$ and so \eqref{eq:Nagumo condition for control2} is satisfied.

\textbf{\RN{7}}: $ \left[ \mu_i = -\gamma \dfrac{\partial \alpha}{\partial \myubi{h}}(q_i), \ \chi_i = 0, \ \psi_i = 0 \right]$. The left-hand-side of \eqref{eq:Nagumo condition for control1} yields $\bar{\eta} - \nu \beta(\myubi{b}) = \bar{\eta} - \nu \beta(\rho) \leq \bar{\eta} - \nu \beta(\ubar{\rho}^\delta)$. From \eqref{eq:eta star} and since $\bar{\eta} \in [0, \eta^*]$, it follows that $\bar{\eta} \leq 2 \bar{\eta}  + \gamma^2 L a \leq \nu \beta(\ubar{\rho}^\delta)$. Thus $\bar{\eta}  - \nu \beta(\ubar{\rho}^\delta) \leq 0$ and \eqref{eq:Nagumo condition for control1} holds.
 
The left-hand-side of \eqref{eq:Nagumo condition for control2} with the substitution of $-\bar{\eta} \geq - 2 \bar{\eta}$  and $\mylbi{b}  = \rho \geq \ubar{\rho}^\delta$  is greater than or equal to the left-hand-side of \eqref{eq:nonconflicting condition1}, and thus Lemma \ref{lem:nonconflicting} ensures \eqref{eq:Nagumo condition for control2} holds.
 
\textbf{\RN{8}}: $ \left[ \mu_i = -\gamma \dfrac{\partial \alpha}{\partial \mylbi{h}}(q_i), \ \chi_i = 0, \ \psi_i = 0 \right]$. The left-hand-side of \eqref{eq:Nagumo condition for control1} with the substitution of $\bar{\eta} \leq 2 \bar{\eta}$ and $\myubi{b} = \rho \geq \ubar{\rho}^\delta$ (see Lemma \ref{lem:b gap}) is less than or equal to the negative of the left-hand-side of \eqref{eq:nonconflicting condition1}, such that \eqref{eq:Nagumo condition for control1} holds via Lemma \ref{lem:nonconflicting}.

The left-hand-side of \eqref{eq:Nagumo condition for control2} yields $-\bar{\eta} + \nu \beta(\myubi{b}) = -\bar{\eta} + \nu \beta(\rho) \geq -\bar{\eta} + \beta(\ubar{\rho}^\delta)$. Again, from \eqref{eq:eta star} and since $\bar{\eta} \in [0, \eta^*]$, it follows that $\bar{\eta} \leq 2 \bar{\eta}  + \gamma^2 L a \leq \nu \beta(\ubar{\rho}^\delta)$. Thus $-\bar{\eta}  + \nu \beta(\ubar{\rho}^\delta) \geq 0$ and \eqref{eq:Nagumo condition for control2} holds.

Finally, since \eqref{eq:Nagumo condition for control1} and \eqref{eq:Nagumo condition for control2} hold for all $i \in \myset{N}_n$, $\tildevar{u} \in \myset{U}$ is a valid control law to enforce \eqref{eq:Nagumos condition matrix} over $\myset{H}^\delta$. This implies that there always exists at least one point-wise solution to $\myvar{u}^*$ from \eqref{eq:consat proposed ct}, namely $\tildevar{u}$. Due to the linearity in the constraints and positive-definiteness of the cost function in \eqref{eq:consat proposed ct}, the solution to $\myvar{u}^*$ is uniquely defined \cite{Nocedal2006}. Thus for any $(\myvar{q}, \myvar{v}) \in\myset{H}^\delta$, there always exists a unique, point-wise solution to \eqref{eq:consat proposed ct}, and  $\myvar{v} \in \myset{V}$.
\end{pf}
\begin{remo}\label{rm:robustness}
Theorem \ref{thm:existenceZCBF} ensures each $b_i$ satisfies the conditions of \rev{Definition \ref{def:zcbf}} on the set $\myset{H}^\delta$ and explicitly uses $\delta$ in the derivation of $\gamma$ and $\nu$. The use of $\delta$ shows how robustness can be incorporated into the control design while respecting input constraints. In the set $\myset{H}^\delta \setminus \myset{H}$, the system \eqref{eq:nonlinear affine dynamics} with \eqref{eq:consat proposed ct} is asymptotically stable to the safe set $\myset{H}$ \cite{Xu2015a}. In other words, for a sufficiently small, bounded perturbation (e.g from model uncertainty) the system will be contained in $\myset{H}^\delta$. 
\end{remo}
The proof of Theorem \ref{thm:existenceZCBF} is constructive and provides insight into designing $\gamma$, $\nu$ to ensure there always exists a solution to \eqref{eq:consat proposed ct}. As discussed in Remark \ref{rm:robustness}, the proposed design considers both constraints on the available control input and robustness with respect to bounded perturbations and sampling time effects. The full ZCBF design is outlined in Algorithm \ref{alg:cbf design}.
\begin{algorithm}
\caption{Control Barrier Function Design}\label{alg:cbf design}
\begin{algorithmic}[1]
\Procedure{ZCBF Design}{$\alpha$, $\beta$, $\delta_0 \geq 0, \bar{\eta}_0 \geq 0$}
\State Determine $\varepsilon$ satisfying Assumption \ref{asm:control} for $\delta_0$, $\bar{\eta}_0$
\State Compute $\gamma_1^*$, $\gamma_2^*$, $\gamma_3^*$ from \eqref{eq:gamma star}, \eqref{eq:gamma star 2}, and \eqref{eq:gamma star 3}.
\State Choose $\gamma \in (0, \min\{ \gamma_1^*,\gamma_2^*, \gamma_3^* \}]$.
\State Compute $\zeta^\delta$, $\ubar{\rho}^\delta$, $\bar{v}$ from \eqref{eq:zeta bar}, \eqref{eq:rho bar}, \eqref{eq:velocity bound}.
\If{ $\delta_0 > 0$}
	\If{ \eqref{eq:delta_star} holds for all $\delta \in [0, \delta_0] $} \State{ Let $\delta^* = \delta_0$  }
	\Else
	 	\State{ Find  $\delta^* \in (0, \delta_0)$ satisfying \eqref{eq:delta_star} }
	 \EndIf	
	 \State Choose $\delta \in (0, \delta^*]$
\ElsIf{ $\delta_0 = 0$}
	\State Find  $\delta^* > 0$ satisfying \eqref{eq:delta_star} 
	\State Set $\delta  = \delta_0 = 0$
\EndIf
\State Compute $\nu_1^*$, $\nu_2^*$ from \eqref{eq:nu star 1}, \eqref{eq:nu star 2} respectively.
\If{ $\delta > 0$ } 
	\State{ Choose $\nu \in [\nu_1^*, \nu_2^*]$ }
\ElsIf{ $\delta = 0$} 
	\State{ Choose $\nu \geq \nu_1^*$ }
	\EndIf 
\State Compute $\eta^*$ from \eqref{eq:eta star}
\If{ $\bar{\eta}_0 = 0 $} 
	\State{Set $\bar{\eta} = 0$} 
\Else
	\State{Choose $\bar{\eta} \in (0, \min \{ \bar{\eta}_0, \eta^*\}]$}
\EndIf
\EndProcedure
\end{algorithmic}
\end{algorithm}
\begin{remo}\label{rm:computation of parameters}
Algorithm \ref{alg:cbf design} presents a guaranteed method of designing ZCBFs for Euler-Lagrange systems with input constraints. The most computationally expensive component involves the computation of $\gamma_2^*$ which requires searching over all $\myvar{q} \in \myset{Q}^\delta$. We note however that the proposed approach requires significantly less computation compared to searching over the entire set $\myset{H}^\delta$. An alternative, albeit more conservative, approach is to bound the terms $\rev{G(\myvar{q})^+}$ and \rev{$\myvar{f}_3(\myvar{q})$} by their respective bounds on \rev{$\myset{H}^\delta$}, as done in \cite{ShawCortez2020}.
\end{remo}
The following corollary ensures the use of Algorithm \ref{alg:cbf design} always ensures a solution to \eqref{eq:consat proposed ct} exists:
\begin{coro}\label{cor:ZCBFalgo}
Consider the system \eqref{eq:nonlinear affine dynamics} with the state and input constraints defined by \eqref{eq:constraint set multiple position}, \eqref{eq:constraint set multiple velocity}, and \eqref{eq:constraint set input}. Given a continuously differentiable extended class-$\mathcal{K}_\infty$ function $\alpha $, extended class-$\mathcal{K}_\infty$ function $\beta$, $\delta_0 \in \mathbb{R}_{\geq 0}$, and $\bar{\eta}_0 \in \mathbb{R}_{\geq 0}$ that satisfy Assumptions \ref{asm:control} and \ref{asm:beta}, Algorithm \ref{alg:cbf design} will always output a $\gamma, \nu \in \mathbb{R}_{>0}$, $\delta, \bar{\eta} \in \mathbb{R}_{\geq 0}$. Additionally if $\delta_0 > 0$, then $\delta$ from Algorithm \ref{alg:cbf design} is strictly positive, and if $\bar{\eta}_0 > 0$, then $\bar{\eta}$ from Algorithm \ref{alg:cbf design} is strictly positive. Furthermore, for this choice of $\alpha$, $\beta$, $\gamma$, $\nu$, $\delta$, and $\bar{\eta}$, let $\myset{H}^\delta$ be defined by \eqref{eq:set H definition delta}. Then there always exists a solution to \eqref{eq:consat proposed ct} for any $(\myvar{q}, \myvar{v}) \in \myset{H}^\delta$.
\end{coro}
\begin{pf}
The proof follows directly from the construction of the ZCBF parameters from Theorem \ref{thm:existenceZCBF}.
\end{pf}

\subsection{Control Implementation}\label{ssec:control implementation}
Theorem \ref{thm:existenceZCBF} ensures the proposed control \eqref{eq:consat proposed ct} is well-posed in that there always exists a unique solution to $\myvar{u}^*$ over $\myset{H}^\delta$. In this section, \rev{we present a sampled-data form of $\myvar{u}^*$ and} ensure forward invariance of $\myset{H}$ of the system \eqref{eq:nonlinear affine dynamics}.

To introduce the sampled-data formulation, we denote $\myvar{q}_k = \myvar{q}(t = t_k)$ and $\myvar{v}_k = \myvar{v}(t = t_k)$ as the sampled states at time $t_k \in \mathbb{R}_{>0}$ for $k \in \mathbb{N}$ and sampling period $T \in \mathbb{R}_{>0}$. To ensure satisfaction of \eqref{eq:Nagumo rel 2 condition} between sampling times\rev{\footnote{Less conservative bounds can be substituted for $\eta(T)$ in this framework so long as the new $\eta(T)$ is of class-$\mathcal{K}$.}}, we formally define $\eta(T)$ as \cite{ShawCortez2019}:
\begin{align}\label{eq:eta sampling}
\eta(T) := \dfrac{(c_1 + c_2 + c_3 c_4) c_5}{c_1 + c_2 c_4}\left( e^{(c_1 + c_2 c_4)T} - 1\right)
\end{align}
where $c_1 \in \mathbb{R}_{>0}$ is the Lipschitz constant associated with the \rev{Lipschitz continuous} function: \rev{$G(\myvar{q})( \myvar{f}_1(\myvar{q}, \myvar{v})+ \myvar{f}_2(\myvar{q}, \myvar{v}) + \myvar{f}_3(\myvar{q}))$}, $c_2 \in \mathbb{R}_{>0}$ is the Lipschitz constant for the extended class-$\mathcal{K}_\infty$ function $\beta$, $c_3 \in \mathbb{R}_{>0}$ is the Lipschitz constant for the \rev{Lipschitz continuous} function $G(\myvar{q})$, $c_4 := \max_{i \in \rev{\myset{N}_m}} u_{max_i}$, and $c_5 := k_{m_\infty}(k_c \bar{v}^2 + k_f \bar{v} + k_g + c_4)$ with $k_{m_\infty} = \max_{\myvar{q} \in \myset{Q}^\delta} \|\rev{G}(\myvar{q}) \|_\infty$, $k_f = \max_{i_\in \rev{\myset{N}_m}} f_i$, \rev{and $k_g = \max_{\myvar{q} \in \myset{Q}^\delta} \|\myvar{f}_3(\myvar{q}) \|_\infty$}.

In regards to the analysis in Section \ref{ssec:analysis}, $\eta(T)$ is substituted for $\bar{\eta}$. In this context, the sampling time $T$ is considered a design parameter and the chosen $\bar{\eta} \in (0,\eta^*]$ defines the maximum allowable sampling frequency for the control law. The use of $\eta(T)$, as explained in \cite{ShawCortez2019}, is to keep the solution $(\myvar{q}(t), \myvar{v}(t))$ ``close enough" to $(\myvar{q}_k, \myvar{v}_k)$ for $t \in [t_k, t_{k+1}]$. This then prevents unsafe behaviour between sampling times. We note that $\eta(T)$ is a class-$\mathcal{K}$ function, which fits with intuition in that as $T$ increases, a larger robustness margin $\eta$ is required to keep the system safe. 

 The proposed sampled-data control law is:
 \begin{align} \label{eq:consat proposed sampled}
\begin{split}
\myvar{u}_k^*(\myvar{q}_k, \myvar{v}_k) \hspace{0.1cm} = \hspace{0.1cm} & \underset{\myvar{u} \in \myset{U}}{\text{argmin}}
\hspace{.3cm} \| \myvar{u} -\myvar{u}_{\text{nom}}(\myvar{q}_k, \myvar{v}_k) \|^2_2  \\
& \text{s.t.} \hspace{.1cm} S \rev{G(\myvar{q}_k) ( \myvar{f}_1(\myvar{q}_k, \myvar{v}_k) +\myvar{f}_2(\myvar{q}_k, \myvar{v}_k)} \\ 
& + \rev{\myvar{f}_3(\myvar{q}_k)} 
+ \myvar{u} ) + \gamma \Lambda(\myvar{q}_k) S \myvar{v}_k \succeq \\
& - \nu \myvar{p}(\myvar{q}_k, \myvar{v}_k) + \eta(T) \myvar{1}_{2n}
\end{split}
\end{align}
Here $\myvar{u}_k^*$ is the ZCBF-based control law which satisfies a sampled-order hold condition between sampling times.

In the following theorem, we ensure safety of the system \eqref{eq:nonlinear affine dynamics} under \eqref{eq:consat proposed sampled}:
\begin{thmo}\label{thm:SafetyControl}
Consider the system \eqref{eq:nonlinear affine dynamics} with the state and input constraint sets defined by \eqref{eq:constraint set multiple position}, \eqref{eq:constraint set multiple velocity}, and \eqref{eq:constraint set input}. Let the sets $\myset{Q}_i^\delta$, $\myset{B}_i^\delta$, and $\myset{H}^\delta_i$ be defined by \eqref{eq:constraint set Ci delta}, \eqref{eq:set B definition delta}, and \eqref{eq:set H definition delta}, respectively, for $i \in \myset{N}_n$ with the continuously differentiable extended class-$\mathcal{K}_\infty$ function $\alpha $ and extended class-$\mathcal{K}_\infty$ function $\beta$. Consider $\gamma_1^*$, $\gamma_2^*$, $\gamma_3^*$, $\nu_1^*$, $\nu_2^*$, $\delta^*$, $\eta^*$ defined, respectively, by \eqref{eq:gamma star 3}, \eqref{eq:gamma star}, \eqref{eq:gamma star 2}, \eqref{eq:nu star 1}, \eqref{eq:nu star 2}, \eqref{eq:delta_star}, \eqref{eq:eta star}. Let $\eta(T)$ be defined by \eqref{eq:eta sampling} for a given sampling time $T \in \mathbb{R}_{> 0}$. Suppose Assumptions \ref{asm:control} and \ref{asm:beta} hold for a sufficiently small $\delta, \bar{\eta} \in \mathbb{R}_{>0}$, and let $\myvar{u}_{nom}: \myset{H}^\delta \to \mathbb{R}^m$ be a given nominal control law. Let $ \delta \in (0, \delta^*]$, $\gamma \in (0, \min\{ \gamma_1^*, \gamma_2^*, \gamma_3^*\}]$, $\nu \in (\nu_1^*, \nu_2^*]$, and further suppose $T$ is small enough such that $\eta(T) \in (0, \bar{\eta}]$ and $\bar{\eta} \leq \eta^*$. Then $\myvar{u}_k^*$ defined by \eqref{eq:consat proposed sampled} exists and is uniquely defined in $\myset{H}^\delta$. Furthermore, if $\beta \circ \myubi{b}$, $\beta\circ \mylbi{b}$ are locally Lipschitz continuous on $\myset{H}_i^\delta$ for all $i \in \myset{N}_n$ and $(\myvar{q}(0), \myvar{v}(0)) \in \myset{H}$, then \eqref{eq:nonlinear affine dynamics} under \eqref{eq:consat proposed sampled} is \textbf{safe}. 
\end{thmo}
\begin{pf}
We note that by Lemma \ref{lem:eta star}, for $\delta \in (0, \delta^*]$, the choice of $\nu$ is well-defined (i.e $(\nu_1^*, \nu_2^*] \neq \emptyset$) and $\eta^*$ is strictly positive. Thus $(0, \eta^*]$ is non-empty and so $\eta(T)$ is well-defined. By Theorem \ref{thm:existenceZCBF}, $\myvar{u}_k^*$ always exists and is uniquely defined on $\myset{H}^\delta$.

\rev{Since the system dynamic terms $\myvar{f}_1$, $\myvar{f}_2$, $\myvar{f}_3$, and $G$ in \eqref{eq:nonlinear affine dynamics} are \rev{globally Lipschitz continuous} and time-invariant, and $\myvar{u}_k$ is a bounded, piece-wise constant function of time, Proposition C.3.7  of \cite{Sontag1998} ensures that an absolutely continuous solution $(\myvar{q}(t), \myvar{v}(t))$ exists for all $t \geq 0$. The conditions of Theorem 3 of \cite{ShawCortez2019} are satisfied for $N = \infty$ such that $\myset{H}$ is forward invariant for all $t \in [0, NT) = [0, \infty)$. Since $\myset{H} \subset \myset{Q} \times \myset{V}$ from Theorem \ref{thm:existenceZCBF}, $(\myvar{q}(t), \myvar{v}(t))$ remains in $\myset{Q} \times \myset{V}$ for all $t \geq 0$ which completes the proof.}
\end{pf}

\rev{A continuous-time version of Theorem \ref{thm:SafetyControl} is presented in the following Corollary:}
\begin{coro}\rev{[Continuous-Time]}\label{cor:SafetyControl ct}
Consider the system \eqref{eq:nonlinear affine dynamics} with the state and input constraint sets defined by \eqref{eq:constraint set multiple position}, \eqref{eq:constraint set multiple velocity}, and \eqref{eq:constraint set input}. Let the sets $\myset{Q}_i^\delta$, $\myset{B}_i^\delta$, and $\myset{H}^\delta_i$ be defined by \eqref{eq:constraint set Ci delta}, \eqref{eq:set B definition delta}, and \eqref{eq:set H definition delta}, respectively, for $i \in \myset{N}_n$ with the continuously differentiable extended class-$\mathcal{K}_\infty$ function $\alpha $ and extended class-$\mathcal{K}_\infty$ function $\beta$. Consider $\gamma_1^*$, $\gamma_2^*$, $\gamma_3^*$, $\nu_1^*$, $\nu_2^*$, $\delta^*$ defined, respectively, by \eqref{eq:gamma star 3}, \eqref{eq:gamma star}, \eqref{eq:gamma star 2}, \eqref{eq:nu star 1}, \eqref{eq:nu star 2}, \eqref{eq:delta_star}. Suppose Assumptions \ref{asm:control} and \ref{asm:beta} hold for a sufficiently small $\delta \in \mathbb{R}_{>0}$ and $\bar{\eta} := 0$, and let $\myvar{u}_{nom}: \myset{H}^\delta \rev{\times \mathbb{R}} \to \mathbb{R}^m$ be a given nominal control law. If $ \delta \in (0, \delta^*]$, $\gamma \in (0, \min \{\gamma_1^*, \gamma_2^*, \gamma_3^*\}]$, $\nu \in [\nu_1^*, \nu_2^*]$, then the control $\myvar{u}^*$ defined by \eqref{eq:consat proposed ct} exists and is uniquely defined on $\myset{H}^\delta$. Furthermore if $\beta \circ \myubi{b}$, $\beta\circ \mylbi{b}$ are locally Lipschitz continuous on $\myset{H}_i^\delta$ for all $i \in \myset{N}_n$, $ \myvar{u}^*$ is locally Lipschitz continuous on $\myset{H}^\delta$, and $(\myvar{q}(0), \myvar{v}(0)) \in \myset{H}$, then \eqref{eq:nonlinear affine dynamics} under \eqref{eq:consat proposed ct} is \textbf{safe}. 
\end{coro}
\begin{pf}
\rev{Due to the Lipschitz properties of the closed-loop system \eqref{eq:nonlinear affine dynamics} under $\myvar{u}^*$ on $\myset{H}^\delta$, for  $(\myvar{q}(0), \myvar{v}(0)) \in \myset{H}$ Theorem 3.1 of \cite{Khalil2002} ensures there exists a time $\delta t_1 \in \mathbb{R}_{>0}$ such that $(\myvar{q}(t), \myvar{v}(t))$ is uniquely defined on $[0, \delta t_1]$. Since the controller $\myvar{u}^*$ enforces the ZCBF conditions of \eqref{eq:Nagumo condition for control1},  Brezis Theorem (Theorem 4 of \cite{Redheffer1972}) ensures that $(\myvar{q}(t), \myvar{v}(t)) \in \myset{B}$ for all $t \in [0, \delta t_1]$. Repeated application of Brezis Theorem ensures then that $\myvar{q}(t) \in \myset{Q}$ on $[0, \delta t_1]$ and so $(\myvar{q}(t), \myvar{v}(t)) \in \myset{H}$ for all $t \in [0, \delta t_1]$. Since $\myset{H}$ is a compact set we can extend the forward invariance interval to $[0, \infty)$ as follows. Since at $\delta t_1$, the state remains in $\myset{H}$, we can repeat the analysis for subsequent times $\delta t_i > \delta t_{i-1}$, $i > 1$ such that $(\myvar{q}(t), \myvar{v}(t)) \in \myset{H}$ for all $t \in [0, \delta t_i]$. To extend $\delta t_i \to \infty$, suppose instead that at some $\bar{t} < \infty$, the state escapes $\myset{H}$. To escape $\myset{H}$, the state must traverse $\myset{H}^\delta \setminus \myset{H}$. However, the control is well-defined on $\myset{H}^\delta$ for which the ZCBF conditions \eqref{eq:Nagumo condition for control1} are always enforced such that the state could never have left $\myset{H}$. This leads to a contradiction wherein no such $\bar{t}$ exists, which leads to forward invariance of $\myset{H}$ for all $t \geq 0$.}
\end{pf}

\begin{remo}[Extension to general constraints]\label{rem:general constraints}
\rev{The generality in the proposed methodology allows for other systems that can be formulated as \eqref{eq:nonlinear affine dynamics} and other nonlinear constraints, which are typical in robot applications (e.g. task-space constraints). To extend to multiple nonlinear constraints, consider a system defined by \eqref{eq:nonlinear affine dynamics} for the generalized coordinates $\tildevar{q}, \tildevar{v}$ and a twice-continuously differentiable constraint function $\myvar{c}(\tildevar{q}): \mathbb{R}^n \to \mathbb{R}^n$ with full rank gradient $\nabla \myvar{c} \in \mathbb{R}^{n \times n}$ and locally Lipschitz Hessian. Now let $\myvar{q} = \myvar{c}(\tildevar{q})$, $\myvar{v} = \nabla \myvar{c}^T \tildevar{v}$ such that $\myvardot{v} = \frac{d}{dt}[\nabla \myvar{c}^T] \tildevar{v} + \nabla \myvar{c}^T \myvardot{v}$. We can now address nonlinear constraints of the form \eqref{eq:constraint set multiple position}, \eqref{eq:constraint set multiple velocity}, and \eqref{eq:constraint set input} for the transformed system. It is straightforward to see that with the full rank assumption that allows for an invertible $\nabla \myvar{c}$ and the bounded Hessian of $\myvar{c}$, the dynamics of the transformed system can be written as \eqref{eq:nonlinear affine dynamics} and still satisfy Properties \ref{prop:M}-\ref{prop:F}. An example of our method for multiple nonlinear constraints will be provided in the next section. In the context of task-space constraint satisfaction for robotics, this transformation requires singular configurations to not be elements of the safe set $\myset{H}^\delta$. However unlike many existing methods, our approach enforces the condition that singular configurations can never be reached instead of simply assuming this to be true.}
\end{remo}

\section{Numerical Examples}\label{sec:examples}
Here we demonstrate the proposed technique in simulation on a 2-DOF planar manipulator. The simulations were performed in Python and the code used for these results along with Algorithm \ref{alg:cbf design} is available at \cite{ShawCortezCode}. We note that the results presented here are accompanied with the corresponding simulation file to recreate the results.

\subsection{Scenario 1}
The manipulator consists of two identical links with a length of $1$ m and mass of $1$ kg, which are parallel to the ground such that $\myvar{g} = 0$. The system is equipped with motors capable of $u_{max_1} = -u_{min_2} = 18$ Nm, and $u_{max_2} = -u_{min_2} = 10$ Nm of torque. The system damping is $F = 0.001 I_{2\times 2}$ kg/s. Let the position/velocity safety constraints be defined by $q_{max_1}= -q_{min_1} = \pi/2$ rad, $q_{max_2}= 5\pi/6$ rad, $q_{min_2} = \pi/2$ rad, and $v_{max_{1,2}} = -v_{min_{1,2}} = 1.5$ rad/s. We choose the following extended class-$\mathcal{K}$ functions for the ZCBFs: $\alpha_1(h) = \tan(h)^{-1}$, $\alpha_2(b) = b^3$. The nominal control is the computed torque control law: $u_{nom} = M(q_2)(\myvarddot{r} - \myvardot{e} - \myvar{e}) + C \myvar{v}$ \cite{Spong1989} where $\myvar{e} = \myvar{q}-\myvar{r}$ and $\myvar{r} = [3.4708 \sin(1.3t), 2.6236 \sin(1.3t)+2.0944]^T$ is the reference that attempts to move the system outside of $\myset{Q} \times \myset{V}$ and $\myset{U}$. This nominal control is used to represent a pre-defined control law or equivalently a human that is incorrectly operating the system. The reader is directed to \cite{ShawCortezCode} for all simulation parameters used.

\begin{figure}[t!]
\centering
	\subcaptionbox{ $q_1(t)$ vs. $t$ \label{fig:q1_sim_1}}
		{\includegraphics[scale=.25]{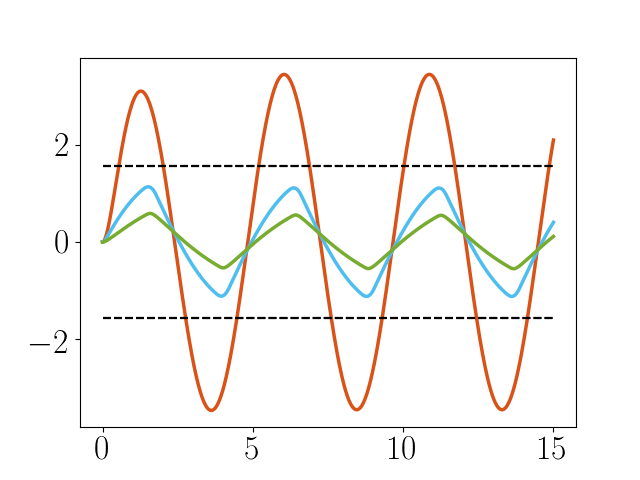}}
	\subcaptionbox{ $q_2(t)$ vs. $t$ \label{fig:q2_sim_1} }
		{\includegraphics[scale=.25]{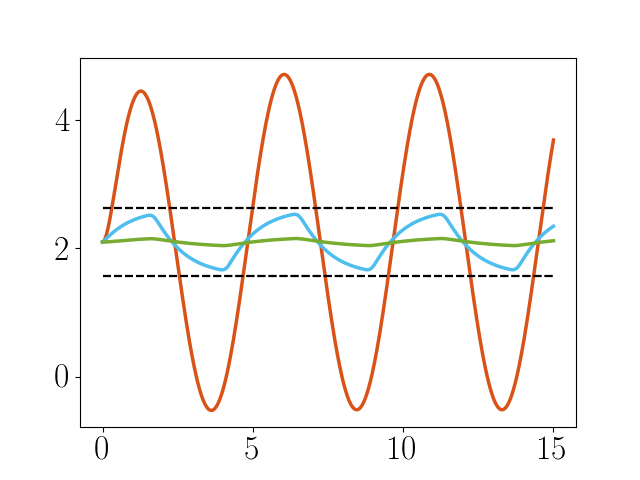}}
	\subcaptionbox{ $v_1(t)$ vs. $t$ \label{fig:v1_sim_1} }
		{\includegraphics[scale=.25]{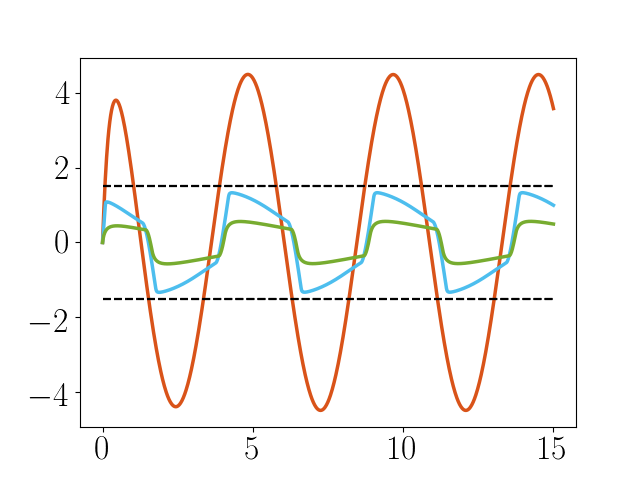}}
	\subcaptionbox{ $v_2(t)$ vs. $t$ \label{fig:v2_sim_1} }
		{\includegraphics[scale=.25]{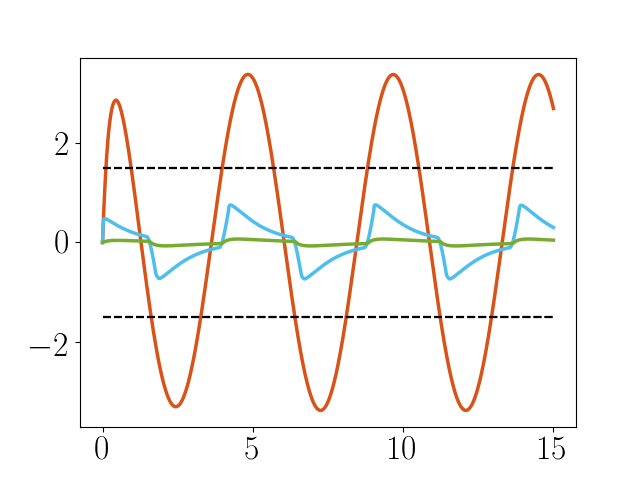}}
	\subcaptionbox{ $u_1(t)$ vs. $t$ \label{fig:u1_sim_1} }
		{\includegraphics[scale=.25]{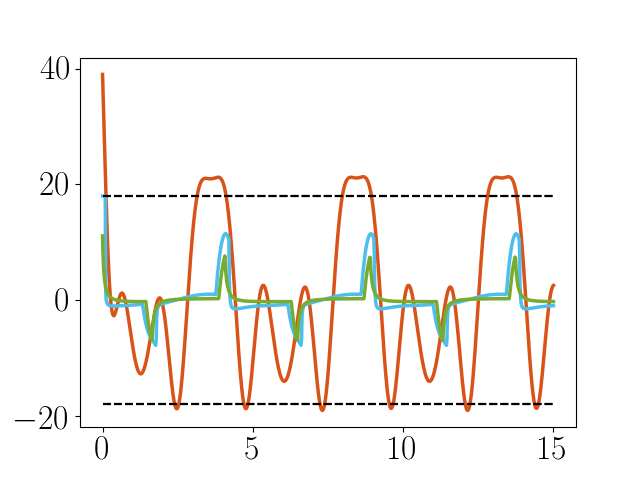}}
	\subcaptionbox{ $u_2(t)$ vs. $t$ \label{fig:u2_sim_1} }
		{\includegraphics[scale=.25]{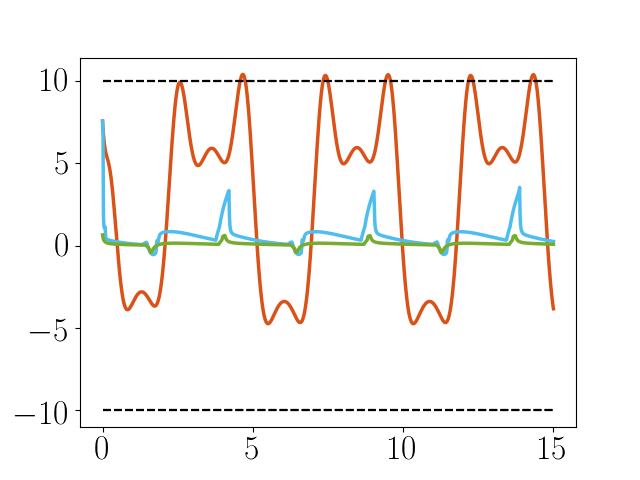}}
	\caption{\rev{(Scenario 1)} Plots of $\myvar{q}$, $\myvar{v}$, and $\myvar{u}$ for the control $\myvar{u} = \myvar{u}_{nom}$ (orange curve), $\myvar{u} = \myvar{u}^*$  from \eqref{eq:consat proposed ct} for the ZCBF parameters from \cite{ShawCortez2020} (green curve), and $\myvar{u} = \myvar{u}^*$  from \eqref{eq:consat proposed ct} for the ZCBF parameters from Algorithm \ref{alg:cbf design} (blue curve). The black-dashed lines depict the boundaries of $\myset{Q}$ in (a), (b), $\myset{V}$ in (c), (d), and $\myset{U}$ in (e), (f), respectively. } \label{fig:cont_time_exp0-2}
\end{figure}

\begin{figure}[t!]
\centering
	\subcaptionbox{ $q_1(t)$ vs. $t$ \label{fig:q1_sim_2}}
		{\includegraphics[scale=.25]{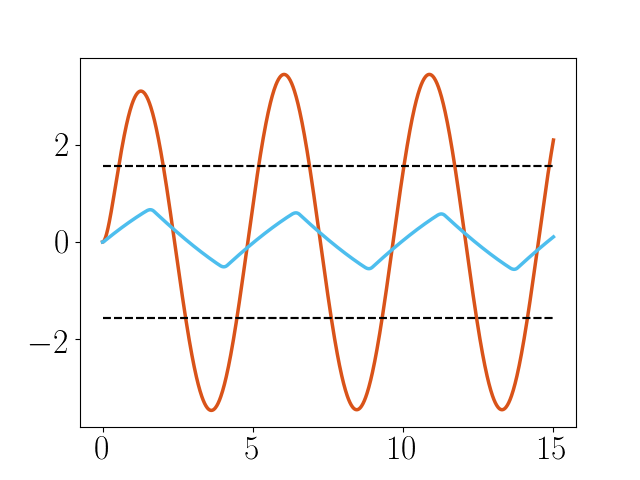}}
	\subcaptionbox{ $q_2(t)$ vs. $t$ \label{fig:q2_sim_2} }
		{\includegraphics[scale=.25]{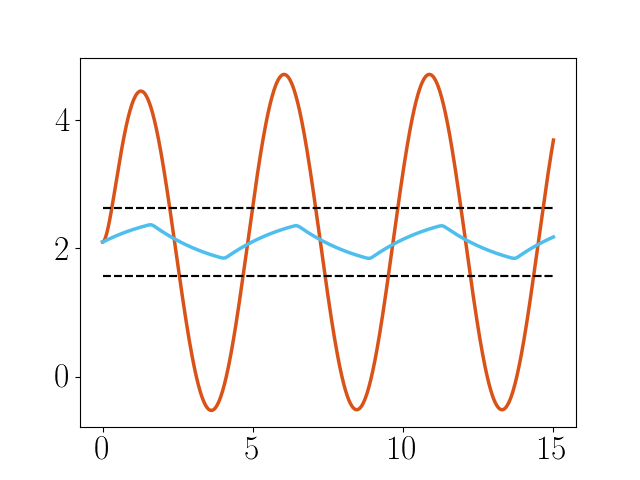}}
	\subcaptionbox{ $v_1(t)$ vs. $t$ \label{fig:v1_sim_2} }
		{\includegraphics[scale=.25]{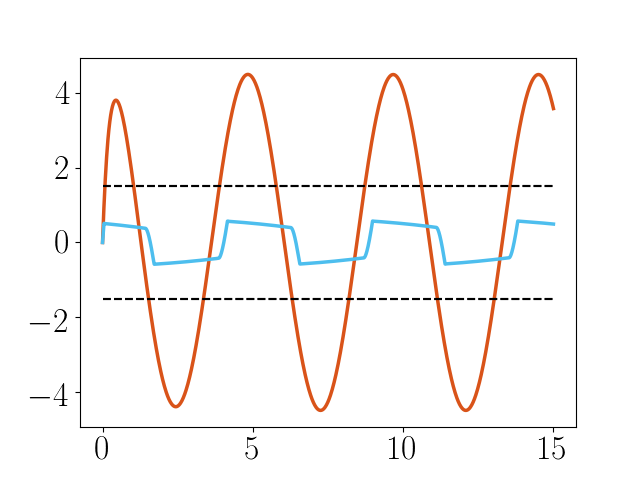}}
	\subcaptionbox{ $v_2(t)$ vs. $t$ \label{fig:v2_sim_2} }
		{\includegraphics[scale=.25]{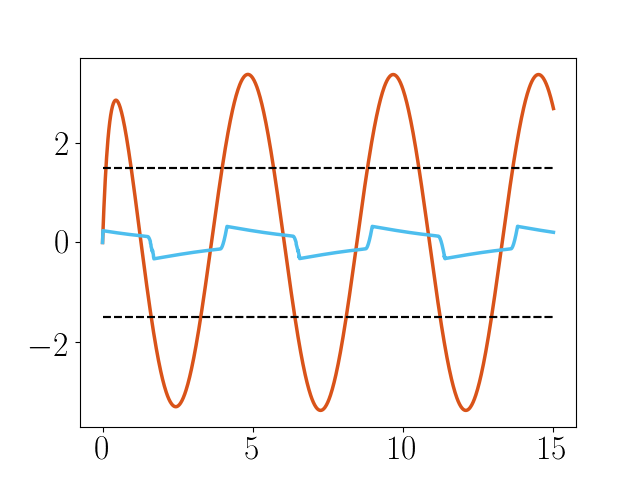}}
	\subcaptionbox{ $u_1(t)$ vs. $t$ \label{fig:u1_sim_2} }
		{\includegraphics[scale=.25]{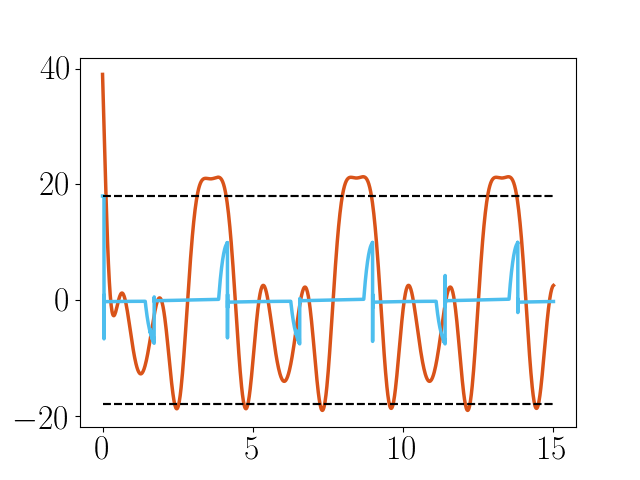}}
	\subcaptionbox{ $u_2(t)$ vs. $t$ \label{fig:u2_sim_2} }
		{\includegraphics[scale=.25]{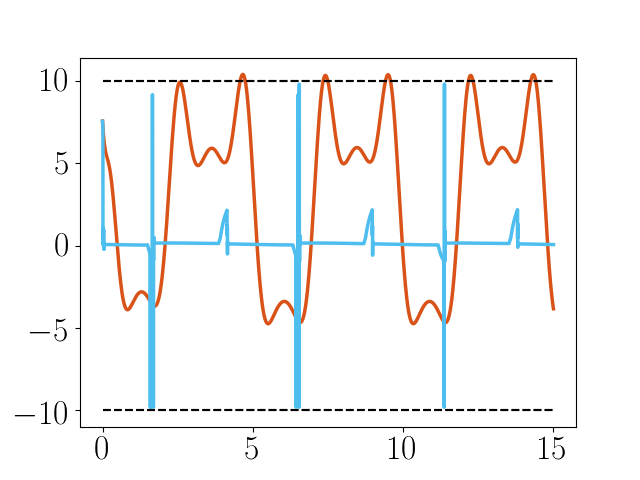}}
	\caption{\rev{(Scenario 1)} Plots of $\myvar{q}$, $\myvar{v}$, and $\myvar{u}$ for the control $\myvar{u} = \myvar{u}_{nom}$ (orange curve) and $\myvar{u} = \myvar{u}_k^*$  from \eqref{eq:consat proposed sampled} for the ZCBF parameters from Algorithm \ref{alg:cbf design} (blue curve). The black-dashed lines depict the boundaries of $\myset{Q}$ in (a), (b), $\myset{V}$ in (c), (d), and $\myset{U}$ in (e), (f), respectively. }\label{fig:cont_time_exp2_discrete}
\end{figure}

First, we compare the proposed technique presented with the preliminary, more conservative method from \cite{ShawCortez2020} in continuous time. Figure \ref{fig:cont_time_exp0-2} shows three system trajectories. The first, depicted in orange, is the system \eqref{eq:nonlinear affine dynamics} subject to the nominal control law, $\myvar{u}_{nom}$, alone. As shown, the nominal control results in violation of all system and input constraints. The second trajectory, depicted in green, shows the result of the system \eqref{eq:nonlinear affine dynamics} subject to the proposed control \eqref{eq:consat proposed ct} (in continuous time) using the ZCBF parameters constructed from \cite{ShawCortez2020} (``ZCBF\_control\_exp1.yaml" from \cite{ShawCortezCode}). The resulting trajectories show satisfaction of all state and input constraints, while attempting to track the nominal control law. This implementation ensures safety, however significant conservativeness is seen by the distances between the trajectories and state/input constraints. The third trajectory, depicted in blue, shows the system \eqref{eq:nonlinear affine dynamics} subject to the proposed control \eqref{eq:consat proposed ct} using the ZCBF parameters constructed from Algorithm \ref{alg:cbf design} (``ZCBF\_control\_exp2.yaml" from \cite{ShawCortezCode}). \rev{The output of Algorithm \ref{alg:cbf design} for the simulations in Figure \ref{fig:cont_time_exp0-2} is:  $\gamma = 1.17$, $\delta = 0.1$, $\nu = 2473.70$,  $\bar{\eta} = 0.0$ for the input parameters: $ \alpha(h) = \tan^{-1}(h)$, $\beta(h) = h^3$, $\delta_0 = 0.1$}. As shown, the controller ensures safety of the overall system, but is also less conservative than the approach from \cite{ShawCortez2020}. One difference between the ZCBF parameter construction between \cite{ShawCortez2020} and Algorithm \ref{alg:cbf design} lies in computation. The method in \cite{ShawCortez2020} only requires the associated bounds from Properties \ref{prop:M}-\ref{prop:F} \rev{and a bound on $\myvar{f}_3$} and scales well with the number of degrees of freedom. Algorithm \ref{alg:cbf design} on the other hand is dependent on searching over some dynamic terms of \eqref{eq:nonlinear affine dynamics} over $\myset{Q}^\delta$. This results in larger computational effort, but yields less conservative behaviour as seen in Figure \ref{fig:cont_time_exp0-2}. By less conservative behaviour, we mean that the state trajectories more closely approach the state constraints for a more aggressive system response.

Next, we note that the results shown in Figure \ref{fig:cont_time_exp0-2} were developed using the continuous time control law \eqref{eq:consat proposed ct}. However, this is dependent on the assumption of local Lipschitz continuity of $\myvar{u}^*$, which is not guaranteed in general. Indeed, under certain parameter configurations (see ``ZCBF\_control\_exp2\_fail.yaml") the system leaves the safe set as a result of discontinuities in the control. When discontinuities occur, $\bar{\eta} > 0$ is required to account for jumps in the control law to ensure forward invariance of the safe set. However, the sampled-data control law \eqref{eq:consat proposed sampled} is able to ensure forward invariance of the safe set for $T = 0.001$ s (see ``ZCBF\_control\_exp2\_discrete.yaml"). The results of the system trajectory subject to the sampled-data controller and ZCBF parameters from Algorithm \ref{alg:cbf design} are shown in Figure \ref{fig:cont_time_exp2_discrete}. \rev{The output of Algorithm \ref{alg:cbf design} for the simulations in Figure \ref{fig:cont_time_exp2_discrete} is: $\gamma = 0.52$, $\delta = 0.01$, $\nu  = 4.57 \times 10^6$, $\bar{\eta} = 6.26$ for the input parameters: $ \alpha(h) = \tan^{-1}(h)$, $\beta(h) = h^3$, $\delta_0 = 0.01$, $\bar{\eta}_0 = 7.0$}.

Figure \ref{fig:cont_time_exp2_discrete} shows the proposed, sampled-data control $\myvar{u}_k^*$ enforcing state constraints, while always respecting input constraints. The effect of incorporating $\bar{\eta}  > 0$ into the control design does impose some conservativeness in the system behaviour. This can be seen by comparing the blue curves between Figures \ref{fig:cont_time_exp0-2} and \ref{fig:cont_time_exp2_discrete}. The state trajectories resulting from the sampled-data control do not approach the state constraints as closely as that of the continuous-time controller. 

\subsection{Scenario 2: Nonlinear Constraints}

\begin{figure}[t!]
\centering
	\subcaptionbox{ $\tilde{q}_2$ vs. $\tilde{q}_1$ \label{fig:sim6_exp5_position}}
		{\includegraphics[scale=.25]{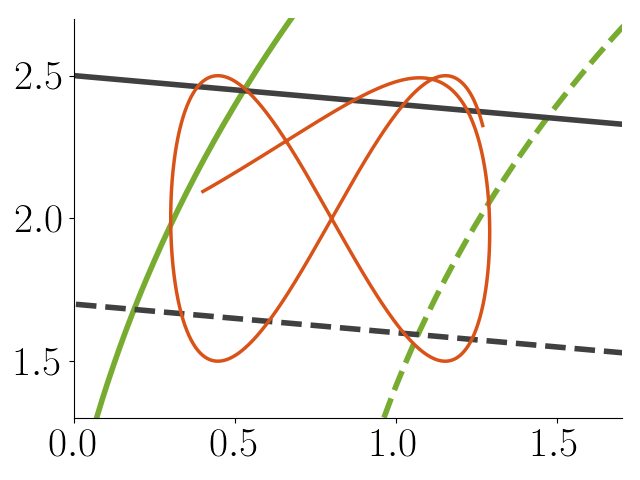}}
	\subcaptionbox{ $\tilde{q}_2$ vs. $\tilde{q}_1$ \label{fig:sim5_exp5_position} }
		{\includegraphics[scale=.25]{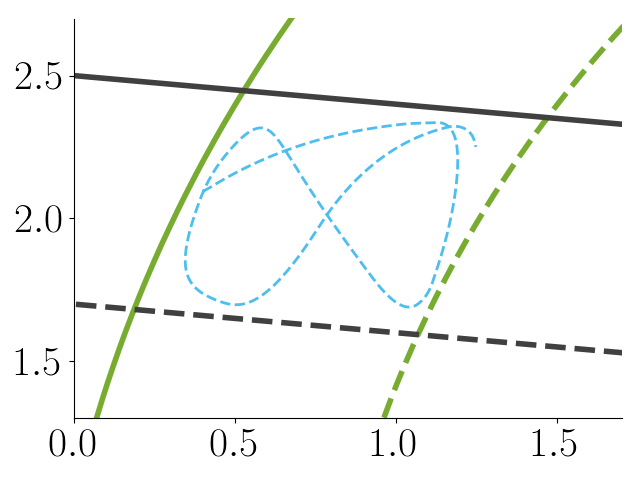}}
	\subcaptionbox{ $v_1(t)$ vs. $t$ \label{fig:sim5_exp5_v0} }
		{\includegraphics[scale=.25]{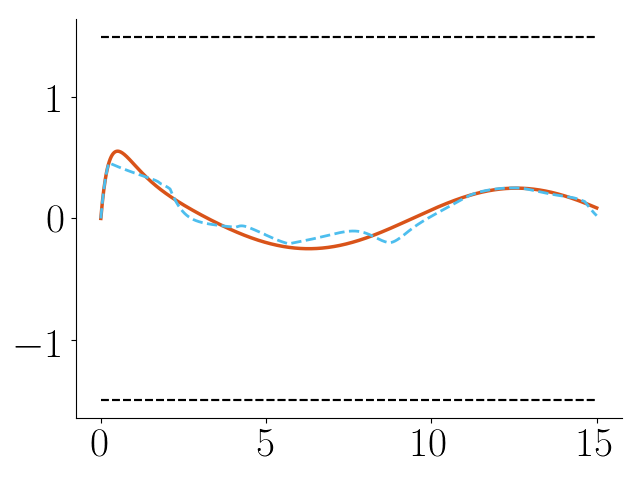}}
	\subcaptionbox{ $v_2(t)$ vs. $t$ \label{fig:sim5_exp5_v1} }
		{\includegraphics[scale=.25]{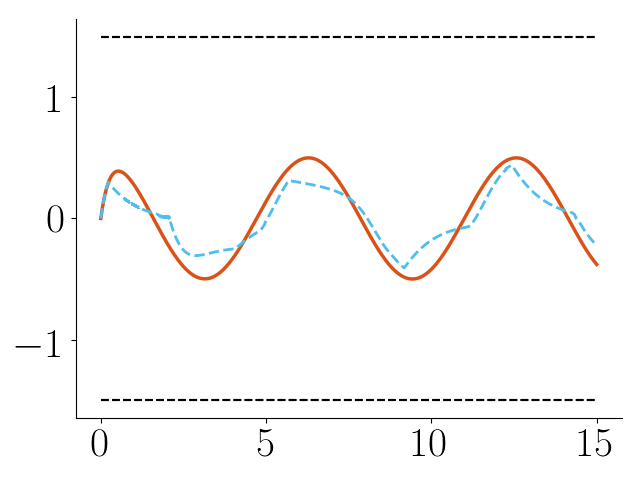}}
	\subcaptionbox{ $u_1(t)$ vs. $t$ \label{fig:sim5_exp5_u0} }
		{\includegraphics[scale=.25]{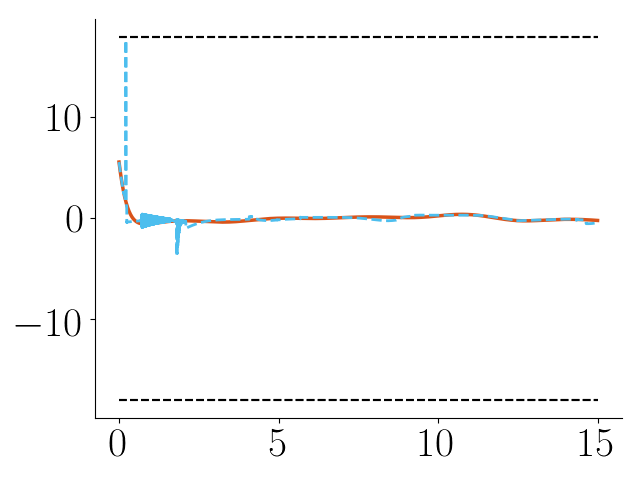}}
	\subcaptionbox{ $u_2(t)$ vs. $t$ \label{fig:sim5_exp5_u1} }
		{\includegraphics[scale=.25]{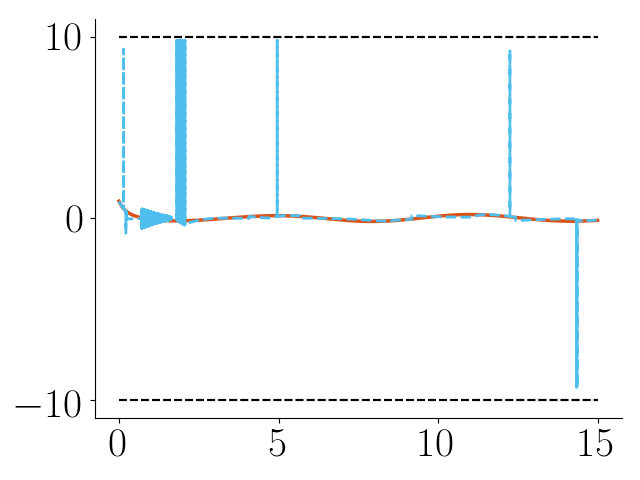}}
	\caption{\rev{(Scenario 2) Plots of $\tildevar{q}$, $\myvar{v}$, and $\myvar{u}$ for the control $\myvar{u} = \myvar{u}_{nom}$ (orange curve) and $\myvar{u} = \myvar{u}_k^*$  from \eqref{eq:consat proposed sampled} for the ZCBF parameters from Algorithm \ref{alg:cbf design} (blue curve). The green and grey curves in (a) and (b) correspond to the ellipsoid and planar constraint level sets outlining $\tilde{\myset{Q}}$, where the solid and dashed curves are associated with $\myvar{q}_{max}$ and $\myvar{q}_{min}$, respectively. The The black-dashed lines depict the boundaries of $\myset{V}$ in (c), (d), and $\myset{U}$ in (e), (f), respectively. }}\label{fig:nonlinear_constraint}
\end{figure}

\rev{We next provide an example of how the proposed methodology can be applied to nonlinear constraints. All model parameters are the same as from Scenario 1 except now the position is bounded by the intersection of ellipsoids and planes, which are presented as follows. We re-define the original joint angles of the 2-DOF manipulator as $\tildevar{q} \in \mathbb{R}^2$. Let $\myvar{c}(\tildevar{q}) = [c_1(\tildevar{q}), c_2(\tildevar{q})]^T$ for $c_1(\tildevar{q}) = -1 + (\tildevar{q} - \myvar{q}_{r_1})^T P (\tildevar{q} - \myvar{q}_r)$, $c_2(\tildevar{q}) = \myvar{q}_{r_2}^T \tildevar{q}$ with $\myvar{q}_{r_1} = [5, 0]^T$, $\myvar{q}_{r_2} = [0.1, 1.0]^T$, and $P = \text{diag}([-1, 0, 0, -1])$. Now we define the (transformed) system state as $\myvar{q} = \myvar{c}(\tildevar{q})$ and define the constraints via \eqref{eq:constraint set multiple position} with $\myvar{q}_{min} = [8, 1.7]^T$ and $\myvar{q}_{max} = [12, 2.5]^T$. A picture of the constraint set $\myset{Q}$ in the joint space, i.e. $\tilde{\myset{Q}}= \{ \tildevar{q} \in \mathbb{R}^2:\myvar{c}(\tildevar{q}) \in \myset{Q} \}$ can be seen in Figure \ref{fig:sim6_exp5_position}. We leave it to the reader to derive the system dynamics using the proposed transformation, but note that $\nabla \myvar{c}$ is full rank for all $\tildevar{q} \in \tilde{\myset{Q}}$. The same velocity and input bounds are used as in Scenario 1. The reference signal for this scenario is: $\myvar{r} = [0.5 \sin(0.5 t) + 0.8, 0.5 \sin(1.0 t)+2]^T$, which yields a figure-eight trajectory (see Figure \ref{fig:sim6_exp5_position}), and all other parameters associated with Scenario 2 can be found under ``ZCBF\_control\_exp5\_nonlinear.yaml" in \cite{ShawCortezCode}.

Algorithm \ref{alg:cbf design} was used to derive the following correct-by-design ZCBF parameters: $\gamma = 1.13$, $\delta = 0.01$, $\nu = 12.7 \times 10^6$, $\bar{\eta} = 7.32$ for the input parameters: $\alpha(h) = \tan^{-1}(h)$, $\beta(h) = h^3$, $\delta_0 = 0.01$, $\bar{\eta}_0 = 8.0$. Figure \ref{fig:nonlinear_constraint} shows a comparison between the nominal tracking controller and the proposed sampled-data controller. As shown in the plots, the proposed control enforces the multiple nonlinear position constraints, while respecting bounds on the (transformed) velocity, and the input constraints simultaneously. The plots show that the proposed control attempts to implement the nominal controller as much as possible, but deviates as necessary to satisfy all the safety constraints.}

Finally, we note some caveats associated with Algorithm \ref{alg:cbf design}. As stated, given any appropriately defined $\alpha$, $\beta$, $\delta_0 \geq 0$, $\bar{\eta}_0 \geq 0$, the algorithm will always output a $\gamma$, $\nu$, and $\bar{\eta}$ such that there exists a $\myvar{u} \in \myset{U}$ to enforce safety. However, the choices of $\alpha$, $\beta$, $\delta_0$, and $\bar{\eta}_0$ are subject to respecting Assumptions \ref{asm:control} and \ref{asm:beta}. Of particular note is Assumption \ref{asm:control} which requires a specified $\varepsilon$ to be known. In general, the choice of ZCBF parameters to ensure $\varepsilon > 0$ is not straightforward. This may result in an iterative procedure to find the appropriate $\alpha$, $\beta$, $\delta_0$, $\bar{\eta}_0$ combination. Furthermore, the use of $T$ as a design parameter may not be representative of real-world systems. Usually a sampling time is \textit{given}. In such a case, iterations over Algorithm \ref{alg:cbf design} will be required to ensure that the appropriate choice of $\alpha$, $\beta$, $\delta_0$, and $\bar{\eta}_0$ yield an $\eta^* \geq \bar{\eta} \geq \eta(T)$. We do note however that the explicit computation of $\eta(T)$ allows for straightforward computation of $\eta^{-1}(\bar{\eta})$ to specify the sampling time required for the given parameters: $\alpha$, $\beta$, $\delta_0$, and $\bar{\eta}_0$, and facilitates the ZCBF design.

\section{Conclusion}
In this paper, we designed multiple, non-conflicting ZCBFs to ensure safety of Euler-Lagrange systems. The design takes into account actuator limitations, robustness margins, and sampling time effects. The proposed design yielded an algorithm to compute safe-by-design ZCBF parameters. \rev{A sampled-data controller was} presented to enforce safety of the Euler-Lagrange system. The proposed approach was demonstrated in simulation on a 2 DOF planar manipulator. Future work will consider simultaneous safety and stability as well as the use of data-based methods to further improve system performance. 

\bibliographystyle{IEEEtran}
\bibliography{IEEEabrv,ShawCortez_CBFConstruction}
\end{document}